\documentclass[aip,jcp,amssymb,reprint,nofootinbib,%
amsmath]{revtex4-1}
\usepackage{hyperref}
\usepackage{epsfig}

\begin{document}
\title{Radiative Corrections to the Off-Resonant Maxwell-Garnett Dielectric Constant and the Spectrum of Emission}
\author{M.~Donaire}
\email{mdonaire@fc.up.pt}
\affiliation{Centro de F\'{\i}sica do Porto, Faculdade de Ci\^{e}ncias da Universidade do Porto, Rua do Campo Alegre 687, 4169-007 Porto, Portugal\\}

\affiliation{Departamento de F\'{\i}sica de la Materia Condensada, Universidad Aut\'{o}noma de
Madrid, E-28049 Madrid, Spain.}

\date{11 April 2010}

\begin{abstract}
We compute the dielectric constant of a Maxwell-Garnett (MG) dielectric renormalized by radiative corrections at off-resonance frequencies. To this aim, the self-energy of the dipole constituents of the dielectric is calculated following Ref.~\onlinecite{PRAvc}. The spectrum of coherent emission is obtained in exact form for the MG model. Contrarily to the common assumption, its expression contains an only local field factor. Beyond the MG model, leading order corrections due to recurrent scattering are implemented. Several experiments are proposed to validate our results in both three and two dimensional samples.
\end{abstract}

\maketitle

\section{Introduction}
The effective dielectric constant in nonpolar fluids, spherical suspensions, metallic nanoparticle composites and generic
disordered media is a problem profusely studied along the history. Its study goes back in time to the seminal works of Claussius \cite{Claussius}, Mossotti \cite{Mossotti},
Lorentz \cite{Lorentz}, Lorenz \cite{Lorenz} and Maxwell-Garnett \cite{Maxwell}. The problem has received the attention of a number of Physics communities.   These include Fluid and Statistical Physics \cite{Yvon,Feldoher,Feldoher2}, Chemical Physics \cite{Kirkwood,Alder1,Feldoher3,Feldoher4,GarciadaBajoReview}, Stellar Atmospheres and Interstellar Dust \cite{Frisch,Purcell,Draine,Draine2},   Multiple-scattering Optics \cite{Foldy,Lax,BulloughHynne,RMPdeVries,Sentenac,Sentenac2}, Quantum Optics and Fluorescence \cite{Alder2,Mukamel,LaudonJPB,LuisRemiMole}. The underlying fundamental theory is that of linear transport in disordered media. Any analytical approach to the problem is based on some model and,  even for the simplest ones, further approximations are needed to achieve meaningful expressions. In all the works cited above the modelling  consists of treating the dielectric as a collection of identical point dipoles which are distributed in space according to some equilibrium statistics. The construction of the dielectric constant is therefore a bottom-up approach in which knowledge about the relevant microscopical details is assumed. The problem reduces this way to one of multiple-dipole interactions. Its solution involves the calculation of a stochastic kernel (in statistical terms), an effective potential (in many-body terminology), an electrical susceptibility (in optics) or a self-energy function (in quantum terminology). To this aim, cluster expansions have been developed \cite{Yvon,BulloughHynne,Feldoher2}.
They all base on perturbative series. Thus, the problem turns into one of finding out the appropriate parameter of the expansion
and of selecting the terms of the resultant series which are relevant in some plausible approximation. The authors of most of those works
justify their analytical approximations with numerical simulations. Numerics have the advantage that one can isolate the physical aspects of
interest by just ignoring those aspects which are not being studied. However, sometimes this leads to oversimplifications which are not physically acceptable.
For instance, the radiative emission of an excited dipole generally depends on its coupling to both electrostatic and radiative modes.
We will argue that those approaches which neglect the contribution of longitudinal modes to radiative emission cannot yield satisfactory results.
On the other hand, those approaches which incorporate only electrostatic interactions cannot be consistent with
the optical theorem. They give incomplete results for the line-width of the spectral function.\\
\indent A more technical aspect is that of the effect of the spatial dispersion of the dielectric function in the renormalization of single-particle polarizabilities. The spatial dispersion  and its effects depend on the specific microscopical structure of the medium. In general, further modelling is needed. Nevertheless, there are still aspects which can be considered generic under a minimum of restrictions. It is the first aim of this article to study the necessary conditions for this to be the case. For this purpose, we analyze the role of the different length scales which enter the problem and identify the appropriate expansion parameters. This allows us to compute, in good approximation, the effective dielectric constant, $\epsilon_{eff}$, of a statistically homogeneous medium off-resonance.  The computation at resonance is left for a separate paper as it involves a different expansion.\\
\indent Closely related to the computation of $\epsilon_{eff}$ is the calculation of the spectrum of stimulated emission. To this aim, we exploit our results in Ref.~\onlinecite{PRAvc}. We address this issue in detail as it serves us to clarify some usual misconceptions regarding the role of local field factors in the spectrum of dipole emission.\\
\indent Our object of study is a complex dielectric medium made of point dipoles embedded in the free space of dielectric constant $\epsilon_{0}$. The latter can be generalized to allow for the embedding in an auxiliary medium  which behaves as a continuum background with well-defined passive constant permitivity. The spatial distribution of dipoles is statistically isotropic and homogeneous. The dipoles polarize the electromagnetic (EM) vacuum. The polarized vacuum  contains additional fluctuations with respect to (\emph{w.r.t.}) the in-free-space vacuum. Those fluctuations are referred to as \emph{polaritons} when coherent. As a result, both the coherent transport features of the medium and the self-energy of the dipole constituents get modified. In turn, that implies that both the dielectric constant and the single-particle polarizabilities, $\alpha$, get renormalized. Our treatment rely on a dipole-Born-Markov  approximation as explained in Ref.~\onlinecite{PRAvc} and is a Green's function based approach.\\
\indent In particular, we deal in this work with a Maxwell-Garnett dielectric. The reasons being that analytical results can be obtained, comparison with previous approaches can be made and experimental tests can be proposed. Generically, we consider a Maxwell-Garnett dielectric as a random medium made of well-separated point dipoles of bare electrostatic polarizability $\alpha_{0}$ whose correlation length $\xi$ satisfies $k\xi\ll1$ for the frequencies of interest, $\omega=ck$. Well-separated means that the distance between dipoles is a few times greater than their longitudinal dimension in order for the dipole approximation to be valid. In the long-wavelength limit of the effective theory, $q\xi\rightarrow0$, no additional assumption about the degree of correlation is needed and the relevant correlation function for long wavelength modes is the two-point function, $h(r)$ \cite{BulloughHynne,Feldoher,vanTigg}. It contains at least two inherent terms. Those are, one which stands for the exclusion volume around each dipole and another one which stands for the self-correlation. Further, if the self-correlation term is discarded, no matter the exact form of the exclusion volume nor the rest of the terms in $h(r)$, the resultant model obeys the usual MG formula for the effective susceptibility \cite{Maxwell,Feldoher,PRLdeVries,Sentenac} (see below).  In this work we will refer to this approximation as MG approximation and to the resultant model which obeys the MG formula as \emph{stricto sensu} (s.s.) MG model. We will denote all the quantities computed in the s.s. MG approximation with the script MG alone.\\
\indent When self-correlation is added to the s.s. MG model, deviations are obtained \cite{Kirkwood,Yvon,Alder1,ElectroFeld}. The resultant theory incorporates recurrent scattering. Differently to the MG model, even in the limit $q\xi\rightarrow0$ no exact solution have been found and further approximations are needed \cite{Feldoher3}.\\
\indent We follow Ref.~\onlinecite{PRAvc} to derive the renormalized expressions of the single-particle polarizability, $\tilde{\alpha}$, and the electrical susceptibility, $\bar{\chi}(q)$.
The addition of radiative corrections amounts to the inclusion of self-polarization effects, closely related
to the concept of local field. Because the self-polarization field depends
itself on the dielectric constant, $\epsilon_{eff}$, the problem becomes one of self-consistency.
 As a result, $\tilde{\alpha}$ is an implicit  function of $\epsilon_{eff}$.\\
\indent Because radiative corrections are affected by the spatial dispersion of $\bar{\chi}(q)$, further approximations are needed. Throughout this work we apply a quasicrystalline approximation \cite{Lax}. It has the convenient property that it is exact for the s.s. MG model in the effective theory limit. In particular, we will apply the s.s. \emph{overlap approximation} \cite{Feldoher} and a variation of it which incorporates recurrent scattering.\\
\indent We will restrict ourselves to purely radiative corrections. Non-radiative effects such as the collisional shift and Doppler and collisional broadening will not be considered. Also, additional quantum correlations between the electrons of neighboring dipoles and the effect of the short-range interaction potentials between dipoles on their individual polarizabilities will be disregarded \cite{Alder1,Alder2}. \\
\indent The paper is organized as follows. In Section \ref{constant} we review our approach in Ref.~\onlinecite{PRAvc} to derive an expression for the renormalized single-particle polarizability. The approximations used in the subsequent sections are described. In Section \ref{lasgamaov} we apply the precedent formalism together with the overlap approximation to the computation of the radiative corrections of the dielectric constant. In Section \ref{MGemi} we compute the spectrum of coherent emission in the three dimensional s.s. MG model. In Section \ref{beyond} we introduce recurrent scattering. We obtain corrections to the s.s. MG dielectric constant and to the spectrum of coherent emission. In Section \ref{Experiments} we propose both numerical and laboratory experiments where our results can be tested. For the measurement of the coherent emission spectrum, formulae in two-dimensional samples are obtained. A numerical example is carried out. In the Discussion section we compare our approach with previous works. Special attention is given to the role of the local field factors.\\
\indent About the notation, we label three-spatial-component vectors with arrows and three-by-three tensors with overlines. We denote the Fourier-transform of
functionals with $q$-dependent arguments instead of the $r$-dependent arguments of their position-space representation.
\section{Bottom-up Approach}\label{constant}
Our procedure to compute the radiative corrections on the dielectric constant involves two complementary steps. In the first one, we renormalize  the single-particle polarizabilities. In the second one, we compute the electrical susceptibility. The former procedure gives rise to the so-called corrections over the Claussius-Mossoti formula \cite{Draine2}. The latter gives rise to the so-called corrections over the Lorentz-Lorenz or Maxwell-Garnett formula \cite{Feldoherp}. The fact that dipoles are treated as point-like allows us to perform both renormalization processes separately. Self-consistency between both of them is required.
\subsection{Renormalization of the single particle porlarizability}
In a statistically homogeneous and isotropic medium, coherent emission of frequency $\omega$  is propagated from an external source by the Dyson propagator, $\bar{G}^{\omega}(q)$. In terms of the susceptibility tensor, $\bar{\chi}^{\omega}(q)$, the transverse and longitudinal components of $\bar{G}^{\omega}(q)$ read,
\begin{eqnarray}
G_{\perp}^{\omega}(q)&=&\frac{1}{k^{2}[1+\chi^{\omega}_{\perp}(q)]-q^{2}},\label{effectivG1}\\
G_{\parallel}^{\omega}(q)&=&\frac{1}{k^{2}[1+\chi^{\omega}_{\parallel}(q)]}\label{effectivG2}.
\end{eqnarray}
The propagator of the self-polarization field was found in Ref.~\onlinecite{PRAvc} to be
\begin{eqnarray}
\mathcal{G}^{\omega}_{\perp}(q)&=&\frac{\chi^{\omega}_{\perp}(q)}{\rho\tilde{\alpha}}G_{\perp}(q)=
\frac{\chi^{\omega}_{\perp}(q)/(\rho\tilde{\alpha})}{k^{2}[1+\chi^{\omega}_{\perp}(q)]-q^{2}},\label{LDOSIper}\\
\mathcal{G}^{\omega}_{\parallel}(q)&=&\frac{\chi_{\parallel}^{\omega}(q)}{\rho\tilde{\alpha}}\:G^{\omega}_{\parallel}(q)
=\frac{1}{\rho\tilde{\alpha}}\:\frac{\chi^{\omega}_{\parallel}(q)}{k^{2}[1+\chi^{\omega}_{\parallel}(q)]}\label{LDOSIparal}.
\end{eqnarray}
In the above formulae, it is implicit that $\chi^{\omega}_{\perp,\parallel}(q)$ can be expanded as a series of one-particle-irreducible (1PI) multiple-scattering terms of order $n$,
\begin{eqnarray}\label{laXenAes}
\chi_{\perp,\parallel}^{\omega}(q)=\sum_{n=1}^{\infty}\chi^{(n)}_{\perp,\parallel}(q,\omega)=\sum_{n=1}^{\infty}X^{(n)}_{\perp,\parallel}(q,\omega)\rho^{n}
\tilde{\alpha}^{n},
\end{eqnarray}
where $\rho$ is the average numerical volume density of scatterers and the functions
$X^{(n)}_{\perp,\parallel}(q,\omega)$  incorporate the spatial dispersion due to the spatial correlations within clusters of $n$ scatterers.
$\bar{\mathcal{G}}^{\omega}(\vec{r},\vec{r})$ is the propagator of virtual photons of frequency $\omega$ emitted and annihilated at the position $\vec{r}$ of a dipole. It amounts to the radiative corrections which give rise to the renormalized polarizability $\tilde{\alpha}$. We define the $\gamma$-factors, $2\gamma_{\perp}^{\omega}$ and $\gamma^{\omega}_{\parallel}$ as the traces of the transverse and longitudinal divergenceless components of $\bar{\mathcal{G}}^{\omega}(\vec{r},\vec{r})$,
\begin{eqnarray}
2\gamma_{\perp}^{\omega}&=&
\int\frac{2\textrm{d}^{3}k}{(2\pi)^{3}}\Bigl[\frac{\chi_{\perp}^{\omega}(q)/(\rho\tilde{\alpha})}{k^{2}
[1+\chi_{\perp}^{\omega}(q)]-q^{2}}-\Re{\{G_{\perp}^{(0)}(q,\omega)\}}\Bigr],
\label{LDOSIperg}\\
\gamma_{\parallel}^{\omega}&=&\int\frac{\textrm{d}^{3}k}{(2\pi)^{3}}\:\Bigl[\frac{1}{\rho\tilde{\alpha}}
\frac{\chi^{\omega}_{\parallel}(q)}{k^{2}[1+\chi^{\omega}_{\parallel}(q)]}-G_{\parallel}^{(0)}(q,\omega)\Bigr],\label{LDOSIparalg}
\end{eqnarray}
where $G_{\perp,\parallel}^{(0)}(q,\omega)$ are the transverse and longitudinal components of the EM field propagator of frequency $\omega$ in free space --i.e., Eqs.(\ref{effectivG1},\ref{effectivG2}) with $\chi^{\omega}_{\perp,\parallel}=0$. In the spatial representation, the longitudinal component corresponds to the electrostatic dipole field propagator,
\begin{equation}\label{stati}
\bar{G}_{stat.}^{(0)}(r,\omega)=\Bigl[\frac{1}{k^{2}}
\vec{\nabla}\otimes\vec{\nabla}\Bigr]\Bigl(\frac{-1}{4\pi\:r}\Bigr),
\end{equation}
while the transverse part corresponds to the radiation field propagator,
\begin{equation}\label{radi}
\bar{G}_{rad.}^{(0)}(r,\omega)=\frac{e^{i\:kr}}{-4\pi
r}\mathbb{I}+\bigl[\frac{1}{k^{2}}\vec{\nabla}\otimes\vec{\nabla}\bigr]\frac{e^{i\:kr}-1}{-4\pi r}.
\end{equation}
Their respective traces contain divergences in the real axis which correspond to the terms subtracted in Eqs.(\ref{LDOSIperg},\ref{LDOSIparalg}).
In Ref.~\onlinecite{PRAvc} we obtained in function of the $\gamma$-factors,
\begin{equation}\label{alpha1}
\tilde{\alpha}(k)=\frac{\alpha_{0}}{1+\frac{2}{3}k^{2}\alpha_{0}\Re{\{\gamma^{(0)}_{\perp}\}}+\frac{1}{3}k^{2}\alpha_{0}
[2\gamma_{\perp}+\gamma_{\parallel}]},
\end{equation}
and parametrized $\tilde{\alpha}$ as a Lorentzian polarizability,
\begin{equation}\label{Lorentzian}
\tilde{\alpha}(k)=\alpha'_{0}k_{res}^{2}[k_{res}^{2}-k^{2}-i\Gamma k^{3}/(c k_{res}^{2})]^{-1}.
\end{equation}
$ck_{res}$, $\Gamma$ and $\alpha_{0}^{'}$ are the renormalized values of the resonant frequency, the line-width and the electrostatic polarizability respectively.
Appropriate regularization of the divergences in terms of the bare resonant frequency, $\omega_{0}=c k_{0}$, and the bare electrostatic polarizability, $\alpha_{0}$, yields \cite{PRLdeVries} $\Re{\{2\gamma_{\perp}^{(0)}\}}=\frac{-3}{k_{0}^{2}\alpha_{0}}$. The renormalized parameters are,
\begin{eqnarray}\label{Gamares}
\Gamma&=&-\frac{c}{3}\alpha'_{0}k^{3}\Im{\{2\gamma^{(0)}_{\perp}+2\gamma_{\perp}+\gamma_{\parallel}\}}|_{k=k_{res}}
\nonumber\\&=&-\Gamma_{0}\frac{2\pi}{k_{0}^{2}}k\Im{\{2\gamma^{(0)}_{\perp}+2\gamma_{\perp}+\gamma_{\parallel}\}}|_{k=k_{res}},
\end{eqnarray}
\begin{equation}\label{alpha0reg}
\textrm{ and }\quad\alpha'_{0}=\alpha_{0}(k_{0}/k_{res})^{2},
\end{equation}
where $\Gamma_{0}=c\alpha_{0}k_{0}^{4}/6\pi$ and $k_{res}$ satisfies,
\begin{equation}\label{kres}
(k/k_{0})^{2}-1=\frac{1}{3}\alpha_{0}k^{2}\Re{\{2\gamma_{\perp}+\gamma_{\parallel}\}}|_{k=k_{res}}.
\end{equation}

\subsection{Correlation function and length-scales}\label{la2point}
As mentioned in the Introduction, spatial dispersion effects depend on the specific microscopical structure of the complex medium. While the dielectric constant $\epsilon_{eff}$ is, by definition, the zero mode of the renormalized dielectric function $\bar{\epsilon}(q)$, the renormalization of the single-particle polarizabilities is affected by near-field effects where the microscopical details matter. As a first approximation, if the medium is not long-range-strongly-correlated, the Kirkwood superposition approximation \cite{Abe} is applicable. That is, in good approximation all the 1PI diagrams can be computed out of the two-point correlation function. Generically, we can approximate
\begin{equation}\label{darriba}
h(r)\simeq-f(r-\xi)+\rho^{-1}\delta^{(3)}(\vec{r})+C\xi\delta^{(1)}(r-\xi).
\end{equation}
In this equation, $f(r-\xi)$ accounts for the exclusion volume around each dipole and its precise form depends on the interaction potential between pairs of scatterers. It tends to one for $r\lesssim\xi$ and to zero for $r\gtrsim\xi$. Usual forms are those of a Lennard-Jones potential and a hard-sphere potential. With no much loss of generality here we will adopt the latter form, $f(r-\xi)=\Theta(r-\xi)$. It behaves as a virtual cavity. The one-dimensional delta function in Eq.(\ref{darriba}) stands for the overdensity of first neighbors around a given dipole. The constant $C$ relates to the coordination number. It might be relevant for high-ordered media. The three-dimensional delta function stands for the self-correlation and it is ignored in the s.s. MG model.\\
\indent Let us set  $C=0$ for the moment and study the relationship between the length-scales of the problem. The purpose of this qualitative analysis is to make an appropriate selection of the relevant diagrams in the cluster expansion. The length-scales are \cite{Mukamel}, the wavelength of the radiation, $\lambda$; the dipole radius, $a$; the radius of the exclusion volume, $\xi$; the average distance between scatterers, $\rho^{-1/3}$; and the effective length associated to the dipole strength, $\sim|\tilde{\alpha}(k)|^{1/3}$. Off-resonance, the latter scale can be identified with that of the bare electrostatic polarizability, $[\alpha_{0}(k)]^{1/3}$. For 'classical dipoles' $\alpha_{0}(k)$ is of the order of $a^{3}$ while for atomic dipoles, $\alpha_{0}(k)\simeq k_{0}^{-2}r_{e}$, $r_{e}$ being the electron radius  \cite{RMPdeVries}. Because the dipole approximation demands $\xi\gtrsim3a$ and $k\xi\ll1$ for an MG dielectric, the neglect of $a$ is generally justified under any circumstance. Hence, dipoles can be treated as Rayleigh scatterers.\\
\indent Off resonance, the scattering cross-section of single dipoles is much smaller than $\xi^{2}$. Recurrent scattering is weak and the most relevant correlation function is that of the exclusion volume, $-\Theta(r-\xi)$. In contrast, at resonance the scattering cross-section of Rayleigh scatterers is of the order of $\lambda^{2}_{res}$. Therefore, there is a change in the scale hierarchy when passing from the off-resonant to the resonant regime which affects the relation between $\xi$ and $|\tilde{\alpha}(k)|^{1/3}$. The influence of the exclusion volume becomes much weaker since the scatterers may overlap optically for $|\tilde{\alpha}(k)|\gtrsim\xi^{3}$. The inherent self-correlation function $\rho^{-1}\delta^{(1)}(\vec{r})$ becomes relevant as it accounts for multiple-reflections between scatterers. That is, recurrent scattering corrections \cite{ElectroFeld,Feldoher3,JPhysCondmat}.\\
\indent In addition, another change in the scale hierarchy might appear at resonance depending on the relation between $\lambda_{res}$ and $\rho^{-1/3}$. In the first place, if $k_{res}\rho^{-1/3}<1$, the convergence of the series in Eq.(\ref{laXenAes}) gets slower. Also, the dipoles overlap optically within a space  where near field effects are dominant. Therefore, in good approximation only the electrostatic dipole-dipole interaction of Eq.(\ref{stati}) is relevant for the recurrent scattering which enters the pair polarizability at resonance. This is confirmed by numerical simulations \cite{JPhysCondmat,Feldoher3}. In the opposite limit, $k_{res}\rho^{-1/3}<1$, far-field interactions must be included based on numerics as well \cite{JPhysCondmat}.\\
\indent The above arguments and the numerical simulations of Ref.~\onlinecite{Alder1,ElectroFeld,Sentenac} make us to conclude that, leaving aside near-field corrections on the single particle polarizabilities, the dielectric constant will be well approximated by the MG formula off-resonance. Also based on the above arguments and the numerical simulations of Ref.~\onlinecite{Mukamel,Feldoher3,Feldoher4}, strong deviations \emph{w.r.t.} MG are expected at resonance. Note that the transition from the off-resonant regime to the resonant one takes place within a very narrow frequency range less than $\Gamma$. That is so since $\Gamma\ll\omega_{res}$ and $k_{res}\xi\ll1$. Therefore, in good approximation, we can treat separately both regimes.\\
\indent In this paper we restrict ourselves to the off resonant regime and treat the recurrent scattering corrections as perturbations. We leave for a separate paper the computation of the line-width and the resonant frequency shift of Eqs.(\ref{Gamares},\ref{kres}) which require calculations at resonance.

\subsection{The quasicrystalline and the overlap approximations}\label{approxis}
\begin{figure}[h]
\includegraphics[height=6.8cm,width=8.8cm,clip]{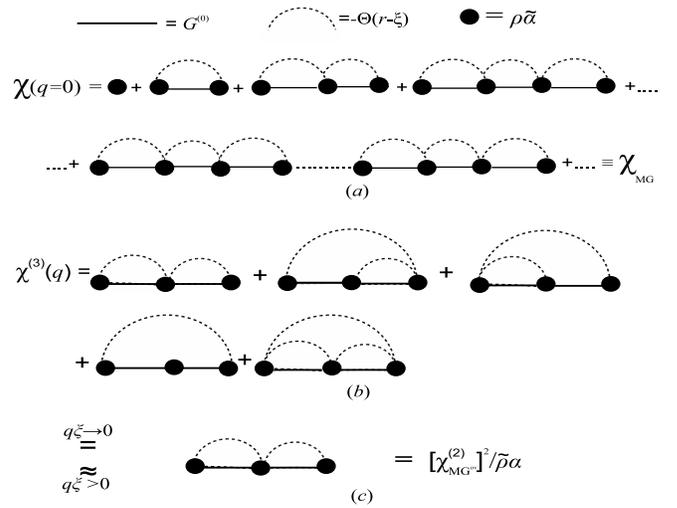}
\caption{($a$) Diagrammatic representation of the geometrical series which amount to $\chi_{MG}$ at $q=0$. ($b$) Series of 1PI diagrams which amount to $\chi^{(3)}$ in the Kirkwood approximation. Only the exclusion volume correlation function enters the MG model. ($c$) In the long-wavelength limit, $q\xi\rightarrow0$, the series in ($b$) is equivalent to the diagram which contains consecutive correlations only \cite{Feldoherp,PRLdeVries}. For shorter wavelengths, ($b$) and ($c$) are related via the overlap approximation \cite{Feldoher}.}\label{MG1}
\end{figure}
It was proved in Ref.~\onlinecite{Feldoher,vanTigg} that, in the long-wavelength limit, $q\xi\rightarrow0$, if only the negative correlation of Eq.(\ref{darriba}) is considered, $h^{MG}(r)=-f(r-\xi)$, the effective susceptibility is exactly that of the MG formula provided that $f(r-\xi)$ is isotropic \cite{Japa},
\begin{equation}
\rho\tilde{\alpha}=\frac{3\chi_{MG}}{(\chi_{MG}+3)}.\label{MGeq}
\end{equation}
The 1PI diagrams which amount to $\bar{\chi}(q)$ are of the kind of those for $\bar{\chi}^{(3)}(q)$ depicted in Fig.\ref{MG1}($b$) in the Kirkwood approximation. In the limits $q\xi\rightarrow0$, $k\xi\rightarrow0$ and for $h^{MG}(r)=-f(r-\xi)$, the series is equivalent to that in Fig.\ref{MG1}($a$) which contains only consecutive correlations.\\
\indent Generically, if that equivalence is extended for any $h(r)$ to shorter wavelengths, the relation becomes an approximation which is referred to as \emph{quasicrystalline approximation} \cite{Lax} (qc). In this approximation, the series of $\bar{\chi}(q,\omega)$ becomes geometrical and the only quantity to be computed is $\bar{\chi}^{(2)}(q,\omega)$,
\begin{equation}\label{laE}
\chi_{\perp,\parallel}^{qc}(q,\omega)=\frac{\rho\tilde{\alpha}}{1-\chi_{\perp,\parallel}^{(2)}(q,\omega)/(\rho\tilde{\alpha})}.
\end{equation}
For the computation of the dielectric constant, only the zero-mode matters, $\chi^{qc}_{eff}(\omega)=\chi^{qc}_{\perp,\parallel}(0,\omega)$. However, spatial dispersion is unavoidable in considering radiative corrections since shorter wavelengths of the above formula are needed for the computation of Eqs.(\ref{LDOSIperg},\ref{LDOSIparalg}).\\
\indent In Ref.~\onlinecite{Feldoher,Feldoherp} the authors worked over a hard-sphere model in the quasicrystalline-MG approximation with $h^{ov}(r)=-\Theta(r-\xi)$. They referred to the resultant approximation as \emph{overlap approximation} --see Fig.\ref{MG1}($c$). In Section \ref{lasgamaov} we apply this approximation to the computation of  radiative corrections. Because self-consistency is demanded there, we will refer to it as \emph{self-consistent overlap MG approximation}.  We will denote the quantities computed in this approximation with a script MG$^{ov}$ and we will use the script MG alone when the computation is independent of the precise form of $f(r-\xi)$. Since it is for $q\xi\rightarrow0$ that the overlap MG approximation becomes exact, the formula
\begin{equation}\label{laEff}
\chi^{\omega}_{MG}=\frac{\rho\tilde{\alpha}}{1-\chi_{\perp,\parallel MG^{ov}}^{(2)}(q=0,\omega)/(\rho\tilde{\alpha})}
\end{equation}
is exact and equals Eq.(\ref{MGeq}).\\
\indent Later on, we add the self-correlation term to $h^{ov}(r)$, $h^{ov}_{rec.}(r)=-\Theta(r-\xi)+\rho^{-1}\delta^{(3)}(\vec{r})$. Its physical effect is the inclusion of corrections due to recurrent scattering (rec.). In this case, the quasicrystalline approximation is not even exact in the long-wavelength limit.  The reason being that the recurrent scattering diagrams involve multiply entangled integrals. Nevertheless, we stick to the quasicrystalline approximation in Section \ref{beyond}. Because still we take $f(r-\xi)=\Theta(r-\xi)$, we will refer to the resultant approximation as \emph{self-consistent overlap recurrent approximation}. We will denote the quantities computed in this approximation with a tilde at the top --except for $\tilde{\alpha}$ in which the tilde denotes the renormalized value of $\alpha$.

\section{Computation of $\epsilon_{eff}$ in the self-consistent overlap MG approximation}\label{lasgamaov}
As mentioned above, in the s.s. MG model the only relevant correlation function is that of the exclusion volume. In such a case Eq.(\ref{MGeq}) is exact and, in combination with Eq.(\ref{alpha1}), we obtain,
\begin{eqnarray}
\epsilon_{MG}^{\omega}&=&1+\frac{\rho\tilde{\alpha}}
{1-\frac{1}{3}\rho\tilde{\alpha}}=1+\chi_{MG}^{\omega}\label{laeps}\\&=&1
+\rho\alpha_{0}k_{0}^{2}
\Bigl[k_{0}^{2}-k^{2}\nonumber\\&+&\frac{1}{3}\alpha_{0}k_{0}^{2}k^{2}(
2\gamma_{\perp}+\gamma_{\parallel})_{k}-\frac{1}{3}\alpha_{0}k_{0}^{2}\rho\Bigr]^{-1}.\nonumber
\end{eqnarray}
In the following, we calculate consistently the $\gamma$-factors in the overlap approximation. The resultant formulae are function of $\xi$ and $\chi_{MG}$. Those formulae together with Eq.(\ref{laeps}) give rise to an implicit function of the complex variable $\chi_{MG}$.\\
\indent Let us compute first $2\gamma_{\perp}$. Applying Eq.(\ref{LDOSIper}) with $\chi_{\perp}(q)$ given by Eq.(\ref{laE}), the transverse polarization propagator reads,
\begin{equation}\label{laG}
\mathcal{G}^{qc}_{\perp}(q)=G_{\perp}^{(0)}(q)
\Bigl[1+\rho\tilde{\alpha}[k^{2}G_{\perp}^{(0)}(q)-\chi^{(2)}_{\perp}(q)/(\rho\tilde{\alpha})^{2}]\Bigr]^{-1}.
\end{equation}
The corresponding $\gamma$-factor is,
\begin{eqnarray}
2\gamma^{qc}_{\perp}&=&\frac{k}{\pi^{2}\zeta}\Bigl\{\int_{0}^{\infty}\textrm{d}Q\frac{Q^{2}}{\zeta^{2}-Q^{2}}
\sum_{n=0}^{\infty}(-\rho\tilde{\alpha})^{n}\label{laGama}\\
&\times&[\frac{\zeta^{2}}{\zeta^{2}-Q^{2}}-(\rho\tilde{\alpha})^{-2}\chi^{(2)}_{\perp}(\zeta,Q)]^{n}\Bigr\}-\Re{\{2\gamma^{(0)}_{\perp}\}},\nonumber
\end{eqnarray}
where $Q\equiv q\xi$. Hereafter we omit the explicit dependence on $\omega=ck$ unless necessary. Using $h^{ov}(r)$ for the calculation of $\chi^{(2)}_{\perp}(\zeta,Q)$ we get,
\begin{eqnarray}
\chi^{(2)}_{\perp MG^{ov}}(q)&=&\frac{(\tilde{\alpha}\rho)^{2}k^{2}}{2}\int\textrm{d}^{3}r e^{i\vec{q}\cdot\vec{r}}\Theta(r-\xi)\nonumber\\&\times&\textrm{Tr}
\{\bar{G}^{(0)}(r)[\mathbb{I}-\hat{q}\otimes\hat{q}]\}\nonumber\\&=&
-\frac{(\tilde{\alpha}\rho)^{2}k^{2}}{2}\int\frac{\textrm{d}^{3}q'}{(2\pi)^{3}}
h^{ov}(|\vec{q}'-\vec{q}|)\Bigl[G_{\perp}^{(0)}(q')\nonumber\\&+&G_{\perp}^{(0)}(q')\cos^{2}{\theta}\:+\:G_{\parallel}^{(0)}(q')\sin^{2}{\theta}\Bigr].
\label{Xiperp}
\end{eqnarray}
where $\hat{q}$ is a unitary vector parallel to $\vec{q}$ and $h^{ov}(Q)=-4\pi\xi^{3}\:j_{1}(Q)/Q=4\pi\xi^{3}\frac{Q\cos{Q}-\sin{Q}}{Q^{3}}$ with $j_{1}$ the spherical Bessel function of first order. At leading order in $\zeta\equiv k\xi$,
\begin{equation}\label{xi2per}
\chi^{(2)}_{\perp MG^{ov}}(Q)\simeq(\rho\tilde{\alpha})^{2}\frac{j_{1}(Q)}{Q}+\mathcal{O}(\zeta^{2}).
\end{equation}
Inserting Eq.(\ref{xi2per}) in Eq.(\ref{laGama}), we can
decompose the resultant formula into long and short wavelength terms up to $\mathcal{O}(\zeta^{0})$,
\begin{equation}
2\gamma_{\perp}^{MG^{ov}}\simeq2\gamma_{\perp MG}^{FF}+2\gamma_{\perp MG^{ov}}^{NN},\qquad\textrm{with}\nonumber
\end{equation}
\begin{eqnarray}
2\gamma^{FF}_{\perp MG}&=&\frac{k}{\pi^{2}\zeta}\Bigl\{
\sum_{n=0}^{\infty}(-\rho\tilde{\alpha})^{n}\int_{0}^{\infty}\textrm{d}Q\frac{Q^{2}}{\zeta^{2}-Q^{2}}\nonumber\\&\times&
[\frac{\zeta^{2}}{\zeta^{2}-Q^{2}}-1/3]^{n}\Bigr\},\label{laGamaCoh1}
\end{eqnarray}
\begin{equation}
2\gamma^{NF}_{\perp MG^{ov}}=\frac{k}{\pi^{2}\zeta}\sum_{n=1}^{\infty}(-\rho\tilde{\alpha})^{n}\int_{0}^{\infty}\textrm{d}Q[j_{1}(Q)/Q]^{n}.\label{laGamainCoh1}
\end{equation}
Eq.(\ref{laGamaCoh1}) carries the far field (FF) propagating modes while Eq.(\ref{laGamainCoh1}) contains the near field (NF) contribution. By taking the limit $\zeta\ll1$, it is immediate to identify in Eq.(\ref{laGamaCoh1}) the term $-1/3\:(\rho\tilde{\alpha})^{2}$ with the zero mode of $-\chi^{(2)}_{\perp MG^{ov}}(Q)$. We take advantage of this to arrange the  $\mathcal{O}(\zeta^{0})$ leading terms of Eq.(\ref{laGamaCoh1}) in an exact expression,
\begin{equation}
2\gamma_{\perp MG}^{FF}(k)=-i\frac{k}{2\pi}\frac{\epsilon_{MG}+2}{3}\sqrt{\epsilon_{MG}}.\label{laGamaA}
\end{equation}
\indent $2\gamma_{\perp MG^{ov}}^{NF}$ amounts to those non-propagating short wavelength modes which scale as $1/\zeta$.
The corresponding series is rapidly convergent. We give below the first five terms of that series,
\begin{eqnarray}\label{laGamaB}
2\gamma_{\perp MG}^{NF}&=&\frac{-k}{2\pi\zeta}\Bigl[\frac{1}{2}\rho\tilde{\alpha}-
\frac{2}{15}(\rho\tilde{\alpha})^{2}+\frac{47}{1280}(\rho\tilde{\alpha})^{3}\label{laGamaB}\\&-&\frac{334}{31185}(\rho\tilde{\alpha})^{4}+
\frac{6891623}{2145927168}(\rho\tilde{\alpha})^{5}+ ...\Bigr]\nonumber.
\end{eqnarray}
\indent Let us consider next the computation of $\gamma_{\parallel}^{MG^{ov}}$. The propagator of longitudinal modes in the quasicrystalline approximation reads,
\begin{equation}\label{laGpar}
\mathcal{G}^{qc}_{\parallel}(q)=\frac{1}{k^{2}}[1+\rho\tilde{\alpha}-\chi_{\parallel}^{(2)}(q)/\rho\tilde{\alpha}]^{-1}.
\end{equation}
In the self-consistent overlap approximation,
\begin{eqnarray}
\chi^{(2)}_{\parallel MG^{ov}}(q)&=&(\rho\tilde{\alpha})^{2}k^{2}\int\textrm{d}^{3}r\:e^{i\vec{q}\cdot\vec{r}}\Theta(r-\xi)\textrm{Tr}
\{\bar{G}^{(0)}(r)[\hat{q}\otimes\hat{q}]\}\nonumber\\&=&-(\rho\tilde{\alpha})^{2}k^{2}\int\frac{\textrm{d}^{3}q'}{(2\pi)^{3}}
h^{ov}(|\vec{q}'-\vec{q}|)\nonumber\\&\times&\Bigl[G_{\parallel}^{(0)}(q')\cos^{2}{\theta}\:+\:G_{\perp}^{(0)}(q')\sin^{2}{\theta}\Bigr].
\label{Xiparall}
\end{eqnarray}
The above equation can be computed in closed form,
\begin{equation}\label{xi2paral}
\chi^{(2)}_{\parallel MG^{ov}}(Q)=(\rho\tilde{\alpha})^{2}[1+\frac{j_{1}(Q)}{Q}f(\zeta)],\quad f(\zeta)=2i\:e^{i\zeta}(i+\zeta).
\end{equation}
Inserting Eq.(\ref{xi2paral}) in Eq.(\ref{laGpar}) and using the MG relationship for $\epsilon_{MG}$ in function of $\rho\tilde{\alpha}$ --Eq.(\ref{laEff}), it is possible to give a closed expression for $\mathcal{G}^{MG^{ov}}_{\parallel}(Q)$,
\begin{equation}\label{lault}
\mathcal{G}^{MG^{ov}}_{\parallel}(Q)=\frac{1}{k^{2}}[1-3\frac{\epsilon_{MG}-1}{\epsilon_{MG}+2}\frac{j_{1}(Q)}{Q}f(\zeta)]^{-1}.
\end{equation}
The use of this formula in the calculation of $\gamma^{MG^{ov}}_{\parallel}$ implies the computation of an infinite series of roots of $1-3\frac{\epsilon_{MG}-1}{\epsilon_{MG}+2}\frac{j_{1}(Q)}{Q}f(\zeta)$ \cite{Feldoher}. Instead of that, we rather expand Eq.(\ref{lault}) in powers of $\rho\tilde{\alpha}$ as this allows us to keep control over the order of accuracy of the series,
\begin{equation}
\mathcal{G}^{MG^{ov}}_{\parallel}(Q)=\frac{1}{k^{2}}\sum_{n=0}^{\infty}(\rho\tilde{\alpha})^{n}(-1)^{n}
f^{n}(\zeta)[j_{1}(Q)/Q]^{n}.
\end{equation}
Finally, we arrive at
\begin{eqnarray}\label{laqtoca}
\gamma^{MG^{ov}}_{\parallel}&=&\frac{k}{2\pi^{2}}\frac{1}{\zeta^{3}}\sum_{n=1}^{\infty}(-1)^{n+1}
(\rho\tilde{\alpha})^{n}f^{n}(\zeta)\nonumber\\&\times&\int\textrm{d}Q\:Q^{2}[j_{1}(Q)/Q]^{n}.
\end{eqnarray}
Note the similarity between the above equation and that for $2\gamma_{\perp MG^{ov}}^{NF}$ in Eq.(\ref{laGamainCoh1}).
$\gamma^{MG^{ov}}_{\parallel}$ contains however additional terms of order $1/\zeta^{3}$ and order zero in $\zeta$ as a result of $f(\zeta)$ in Eq.(\ref{xi2paral}). We give below the first five terms of the orders $\mathcal{O}(\zeta^{-3})$, $\mathcal{O}(\zeta^{-1})$ and $\mathcal{O}(\zeta^{0})$,
\begin{widetext}
\begin{eqnarray}
\gamma^{MG^{ov}}_{\parallel}&\simeq&\frac{-k}{2\pi}\frac{1}{\zeta^{3}}[\rho\tilde{\alpha}+\frac{2}{3}(\rho\tilde{\alpha})^{2}
+\frac{5}{24}(\rho\tilde{\alpha})^{3}+\frac{272}{2835}(\rho\tilde{\alpha})^{4}+\frac{40949}{870912}(\rho\tilde{\alpha})^{5}+...]\label{m1}\\
&-&\frac{k}{2\pi}\frac{1}{\zeta}[\frac{1}{2}\rho\tilde{\alpha}+\frac{2}{3}(\rho\tilde{\alpha})^{2}
+\frac{5}{16}(\rho\tilde{\alpha})^{3}+\frac{544}{2835}(\rho\tilde{\alpha})^{4}+\frac{204745}{1741824}(\rho\tilde{\alpha})^{5}+...]\label{m2}\\
&-&i\frac{k}{2\pi}[\frac{1}{3}\rho\tilde{\alpha}+\frac{4}{9}(\rho\tilde{\alpha})^{2}
+\frac{5}{24}(\rho\tilde{\alpha})^{3}+\frac{1088}{8505}(\rho\tilde{\alpha})^{4}+\frac{204745}{2612736}(\rho\tilde{\alpha})^{5}+...]\label{m3}.
\end{eqnarray}
\end{widetext}
\indent Because $2\gamma^{FF}_{\perp MG}$ contains only long wavelength modes, the overlap approximation is there as good as for the computation of $\chi_{MG}$ at order $\mathcal{O}(\zeta^{0})$. Hence, the superscript MG instead of MG$^{ov}$ in Eq.(\ref{laGamaA}). On the contrary, $\gamma^{MG^{ov}}_{\parallel}$  together with $2\gamma^{NF}_{\perp MG^{ov}}$ contain the contribution of short wavelength photons which probe the microscopic structure of the dielectric. The application of the overlap approximation in them is only exact for the $\mathcal{O}(\rho\tilde{\alpha})$ term. In function of $\chi_{MG}$, the $\gamma$-factors read,
\begin{widetext}
\begin{eqnarray}
2\gamma_{\perp}^{MG^{ov}}+\gamma^{MG^{ov}}_{\parallel}
&=&-i\frac{k}{2\pi}\frac{\epsilon_{MG}+2}{3}\sqrt{\epsilon_{MG}}\label{g1}\\
&-&\frac{i k}{2\pi}[\frac{1}{3}\chi_{MG}
+\frac{1}{3}\chi^{2}_{MG}-0.051\chi^{3}_{MG}+0.055\chi^{4}_{MG}-0.015\:\chi^{5}_{MG}+...]\label{g2}\\
&-&\frac{k}{2\pi}\frac{1}{\zeta}[\chi_{MG}
+\frac{1}{5}\chi^{2}_{MG}+0.105\chi^{3}_{MG}-0.027\chi^{4}_{MG}+0.006\:\chi^{5}_{MG}+...]\label{g3}\\
&-&\frac{k}{2\pi}\frac{1}{\zeta^{3}}[\chi_{MG}
+\frac{1}{3}\chi^{2}_{MG}-\frac{1}{8}\chi^{3}_{MG}+0.07\:\chi^{4}_{MG}-0.03\:\chi^{5}_{MG}...]\label{g4}.
\end{eqnarray}
\end{widetext}
As advanced, the $\gamma$-factors in Eq.(\ref{g1}) are function of $\xi$ and $\chi_{MG}$. Because $\chi_{MG}$ is also a function of the $\gamma$-factors in Eq.(\ref{laeps}), the computation of $\chi_{MG}$ becomes a problem of self-consistency.

\section{The Spectrum of MG Polaritons and Coherent Emission}\label{MGemi}
It was suggested in Ref.~\onlinecite{PRAvc} that the coherent radiation emitted by one of the dipoles in a complex medium could be collected in the far field. According to the standard Beer-Lambert law,
\begin{equation}\label{WCohprop}
W^{Coh.}(|\vec{r'}-\vec{r}|)=W^{Coh.}(\vec{r})\exp{[-2k|\vec{r'}-\vec{r}|\kappa]},
\end{equation}
where $\kappa$ is the extinction coefficient, $\kappa=\Im{\{\sqrt{\epsilon_{MG}}\}}$ and $W^{Coh.}(\vec{r})$ is the coherent power radiated from the emitter site, $\vec{r}$. For a statistically homogenous medium,
\begin{equation}\label{WCOHA}
W^{Coh.}(\vec{r})=W_{o}\int\frac{\textrm{d}^{3}q}{(2\pi)^{3}}\:2\Re{\{\frac{\chi_{\perp}(q)}{\rho\tilde{\alpha}}\}}\Im{\{G_{\perp}(q)\}},
\end{equation}
where $W_{o}=\frac{-\omega^{3}}{6c^{2}\epsilon_{0}}|\vec{p}_{0}|^{2}$, being $\vec{p}_{0}=\epsilon_{0}\tilde{\alpha}\vec{E}^{\omega}_{0}(\vec{r})$ the dipole moment induced on the emitter by an external field of frequency $\omega$, $\vec{E}^{\omega}_{0}(\vec{r})$. The integral of Eq.(\ref{WCOHA}) is, up to prefactors, the spectral function of transverse polaritons,
\begin{equation}
\textrm{LDOS}^{Pols.}_{\perp}(\omega)=-\frac{\omega}{\pi c}\int\frac{\textrm{d}^{3}q}{(2\pi)^{3}}\Re{\{\frac{\chi_{\perp}(q)}{\rho\tilde{\alpha}}\}}\Im{\{G_{\perp}(q)\}}.\label{LDOSsl}
\end{equation}
The poles of $\Im{\{G_{\perp}(q)\}}$, $k_{\perp}^{Pol.}$, satisfy the dispersion relations for transverse normal modes, $k^{2}\epsilon_{\perp}(q)-q^{2}|_{q=k^{Pol.}_{\perp}}=0$. The residue $\Re{\{\frac{\chi_{\perp}(k_{\perp}^{Pol.})}{\rho\tilde{\alpha}}\}}$ was interpreted in Ref.~\onlinecite{PRAvc} as a renormalization factor. For an MG dielectric, the condition $k\xi\ll1$ precludes the existence of geometrical resonances and $G_{\perp}(q)$ contains a simple pole at $k_{\perp}^{Pol.}=\sqrt{\epsilon_{MG}}$. In the s.s. MG model, $\Re{\{\frac{\chi_{\perp}(k_{\perp}^{Pol.})}{\rho\tilde{\alpha}}\}}=\frac{\Re{\{\epsilon_{MG}\}}+2}{3}$, which is a Lorentz-Lorenz local field factor. Thus, Eq.(\ref{LDOSsl}) reads,
\begin{equation}
\textrm{LDOS}^{Pols.}_{MG\perp}(\omega)=\frac{\omega^{2}}{4\pi^{2} c^{2}}\:\frac{\Re{\{\epsilon_{MG}\}}+2}{3}\Re{\{\sqrt{\epsilon_{MG}}\}}.\label{LDOSMG}
\end{equation}
\indent We do not address here the longitudinal polaritons as they require further discussion \cite{Bullough}. They are irrelevant in the present context.\\
\indent Making use of the MG formula we can write $\vec{p}_{0}$ as an effective spherical dipole of volume $\rho^{-1}$ and dielectric constant $\epsilon_{MG}$, $\vec{p}_{0}=3\epsilon_{0}\rho^{-1}\frac{\epsilon_{MG}-1}{\epsilon_{MG}+2}\vec{E}^{\omega}_{0}(\vec{r})$. Thus, we obtain,
\begin{equation}\label{wcohMG}
W^{Coh.}_{MG}(\vec{r})=\mathcal{W}^{0}_{\omega}\:\Bigl|3\frac{\epsilon_{MG}-1}{\epsilon_{MG}+2}\Bigr|^{2}\:\frac{\Re{\{\epsilon_{MG}\}}+2}{3}\Re{\{\sqrt{\epsilon_{MG}}\}},
\end{equation}
where $\mathcal{W}_{\omega}^{0}=\frac{\epsilon_{0}\omega^{4}}{12\pi c^{3}\rho^{2}}|\vec{E}^{\omega}_{0}(\vec{r})|^{2}$.
Note that Eqs.(\ref{LDOSMG},\ref{wcohMG}) are not subject to the overlap approximation but just to the MG approximation.
\section{Beyond the MG model}\label{beyond}
In this Section we correct the MG and MG$^{ov}$ formulae with recurrent scattering terms due to the
addition of the self-correlation which is ignored in the overlap MG approximation.
\subsection{Off-resonant recurrent scattering}
\begin{figure}[h]
\includegraphics[height=3.25cm,width=7.0cm,clip]{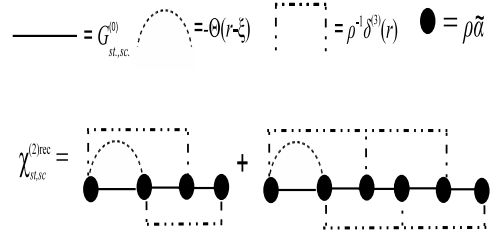}
\caption{Diagrammatic representation of the two-scatterer recurrent susceptibilities of Eq.(\ref{Xirecs}).}\label{MG3}
\end{figure}

The overlap recurrent approximation consists of using $h^{ov}_{rec}(r)=-\Theta(r-\xi)+\rho^{-1}\delta^{(3)}(\vec{r})$ and applying the quasicrystalline approximation. The variation of the effective MG susceptibility due to the introduction of recurrent scattering in the long-wavelength limit has been studied by a number of authors \cite{Kirkwood,Yvon,ElectroFeld,Alder1}. We include short-wavelength variations. It is worth noting here a subtle point. Even though the recurrent scattering  processes (denoted with the script rec.) are formally equivalent to some of the virtual self-polarization diagrams which correct the bare polarizability, the physical content is different. That is so because the recurrent processes amount to the multiple reflections that actual probe photons, instead of virtual, experience within clusters of scatterers.\\
\indent The two-scatterer diagrams of $\bar{\tilde{\chi}}^{(2)rec.}_{st.,sc.}$ are depicted in Fig.\ref{MG3}. They include an infinite number of reflections between two particles. As it was explained in Section \ref{la2point}, we will consider separately the purely electrostatic (near field) contribution and the purely far field (scalar-like) contribution. In these two limits, we have the respective simplified propagators,
\begin{eqnarray}
\bar{G}^{(0)}_{st.}(\vec{r})&=&(1-3\hat{r}\hat{r})/4\pi k^{2}r^{3},\label{estato}\\
\bar{G}^{(0)}_{sc.}(r)&=&[-e^{ikr}/4\pi r]\mathbb{I}.\label{scalar}
\end{eqnarray}
The corresponding 'two-scatter recurrent susceptibilities' read,
\begin{eqnarray}
\bar{\tilde{\chi}}^{(2)rec.}_{st.,sc.}(\vec{r},\omega)&=&-\frac{(\rho\tilde{\alpha})^{2}}{k^{2}}\bar{G}^{(0)}_{st.,sc.}(\vec{r})\label{Xirecs}\\
&\cdot&\sum_{m=1}(k^{2}\tilde{\alpha})^{2m}
[\bar{G}^{(0)}_{st.,sc.}(\vec{r})]^{2m}\:[1-\Theta(r-\xi)].\nonumber
\end{eqnarray}
For the electrostatic contribution,
\begin{eqnarray}
\bar{\tilde{\chi}}^{(2)rec.}_{st.}(\vec{r},\omega)&=&\frac{(\rho\tilde{\alpha})^{2}}{4\pi}[1-\Theta(r-\xi)]\frac{-(\tilde{\alpha}/4\pi)^{2}}{r^{3}(r^{6}-(\tilde{\alpha}/4\pi)^{2})}
\nonumber\\&\times&
\Bigl[\mathbb{I}\:-\:3\hat{r}\hat{r}\frac{3r^{6}-4(\tilde{\alpha}/4\pi)^{2}}{r^{6}-4(\tilde{\alpha}/4\pi)^{2}}\Bigr].\label{recur}
\end{eqnarray}
The zero-mode of the above formula yields a correction over the factor $1/3(\rho\tilde{\alpha})^{2}$ of the usual MG formula. At leading order,
\begin{equation}
\tilde{\chi}_{\perp,\parallel st.}^{(2)rec.}(Q=0)\simeq\frac{1}{3}(\rho\tilde{\alpha})^{2}(\tilde{\alpha}/4\pi\xi^{3})^{2}\label{xieffrec}.
\end{equation}
As expected, the longitudinal and transverse zero modes are equal.  Beyond the long-wavelength limit,
\begin{eqnarray}
\tilde{\chi}_{\perp st.}^{(2)rec.}(Q)&\simeq&(\rho\tilde{\alpha})^{2}\frac{j_{1}[Q]}{Q}(\tilde{\alpha}/4\pi\xi^{3})^{2},\label{xiperprec}\\
\tilde{\chi}_{\parallel st.}^{(2)rec.}(Q)&\simeq&-2\chi_{\perp st.}^{(2)rec.}(Q)+(\rho\tilde{\alpha})^{2}\frac{1}{120}
(\tilde{\alpha}/4\pi\xi^{3})^{2}\label{xiparalrec}\\&\times&\Bigl[\cos{Q}(120-6Q^{2}+Q^{4})+\frac{\sin{Q}}{Q}(720\nonumber\\&-&24Q^{2}+2Q^{4}-Q^{6})
+Q^{6}\textrm{CosIntegral}Q\Bigr].\nonumber
\end{eqnarray}
We find that, off-resonance, the dimensionless expansion parameter for the electrostatic recurrent susceptibility is $\tilde{\alpha}/4\pi\xi^{3}$. This was already noticed by Kirkwood and Yvon \cite{Kirkwood,Yvon}. The precise computation depends on the type of interaction between the scatterers. For a hard-sphere potential Alder \emph{et al.} \cite{Alder1}  and Cichocki and Felderhof \cite{ElectroFeld} computed numerically $\bar{\chi}^{rec.}_{st.}(Q=0)$ as a series expansion on $z\equiv2^{3}\tilde{\alpha}/4\pi\xi^{3}$. They found for the $\mathcal{O}(z^{2})$ term a numerical factor in the range $[0.016,0.028]$ which depends slightly on the filling factor. Our analytical calculation yields $1/2^{6}\simeq0.016$.\\
\indent For the far-field scalar-like contribution,
\begin{eqnarray}
\bar{\tilde{\chi}}^{(2)rec.}_{sc.}(\vec{r},\omega)&=&-(\rho\tilde{\alpha})^{2}k^{6}\tilde{\alpha}^{2}[1-\Theta(r-\xi)]\nonumber\\
&\times&\frac{[\bar{G}^{(0)}_{sc.}(r)]^{3}}{1-k^{4}\tilde{\alpha}^{2}[\bar{G}^{(0)}_{sc.}(r)]^{2}}\mathbb{I}.
\end{eqnarray}
Its zero-mode at leading order is,
\begin{equation}
\tilde{\chi}^{(2)rec.}_{sc.}(Q=0)\simeq-(\rho\tilde{\alpha})^{2}\:k^{6}\tilde{\alpha}^{2}/4\pi[\gamma_{Eu}-1+\lg{3\zeta}-i\pi/2].
\end{equation}
Beyond the long-wavelength limit,
\begin{eqnarray}
\tilde{\chi}^{(2)rec.}_{sc.}(Q)&\simeq&\frac{1}{2}(\rho\tilde{\alpha})^{2}\:k^{6}\tilde{\alpha}^{2}/4\pi\Bigl[2(1-\gamma_{Eu})-\lg{(9\zeta^{2}-Q^{2})}\nonumber\\
&+&\frac{3\zeta}{Q}\lg{(\frac{3\zeta-Q}{3\zeta+Q})}+i\pi(3\zeta/Q+|1-3\zeta/Q|)\Bigr],\nonumber
\end{eqnarray}
where $\gamma_{Eu}$ is the Euler constant.
We find that the dimensionless expansion parameter for the far-field recurrent susceptibility is $\tilde{\alpha}k^{3}$. This parameter is much smaller than $\tilde{\alpha}/4\pi\xi^{3}$ in an MG dielectric. Therefore, the discard of the scalar contribution is justified in good approximation.

\subsection{Computation of $\epsilon_{eff}$ in the self-consistent overlap recurrent approximation}

\begin{figure}[h]
\includegraphics[height=3.6cm,width=7.0cm,clip]{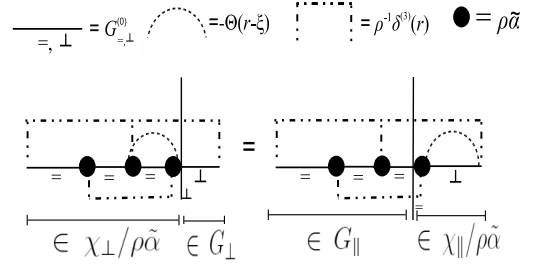}
\caption{Diagrammatic representation of the equivalence between the transverse $\tilde{\gamma}$-terms (l.h.s. diagram) and the longitudinal $\tilde{\gamma}$-term (r.h.s.) of orders $\mathcal{O}(\zeta^{-1})$ and $\mathcal{O}(\zeta^{0})$.}\label{MG4}
\end{figure}

We proceed to compute the modified $\gamma$-factors. To do so, we include the recurrent terms of Eqs.(\ref{xieffrec}-\ref{xiparalrec}) in $\bar{\tilde{\chi}}^{(2)}(Q,\omega)$,
\begin{equation}
\bar{\tilde{\chi}}^{(2)}(Q,\omega)=\bar{\chi}^{(2)}_{MG^{ov}}(Q,\omega)+\bar{\tilde{\chi}}^{(2)rec.}_{st.}(Q,\omega),
\end{equation}
and apply the quasicrystalline approximation which leads to $\bar{\tilde{\chi}}(Q,\omega)$ via  Eq.(\ref{laE}). By comparing Eqs.(\ref{xieffrec}-\ref{xiparalrec}) with those corresponding to the overlap MG approximation, the computation of $\bar{\tilde{\chi}}^{(2)}(Q,\omega)$ just involves the substitution of the factor $(\rho\tilde{\alpha})^{2}$ by $(\rho\tilde{\alpha})^{2}[1+(\tilde{\alpha}/4\pi\xi^{3})^{2}]$ in $\bar{\chi}^{(2)}_{MG^{ov}}(Q,\omega)$. The resultant formulae are exact, in our approximation, up to $\mathcal{O}(\tilde{\alpha}^{4})$ in $\tilde{\chi}_{eff}$ and up to $\mathcal{O}(\tilde{\alpha}^{3})$ in the $\tilde{\gamma}$-factors. Higher order accuracy can be obtained if needed out of Eq.(\ref{recur}). The relation between $\rho\tilde{\alpha}$ and $\tilde{\chi}_{eff}$ gets modified as
\begin{equation}\label{tila}
\tilde{\chi}_{eff}=\rho\tilde{\alpha}\Bigl[1-\frac{1}{3}\rho\tilde{\alpha}[1+(\tilde{\alpha}/4\pi\xi^{3})^{2}]\Bigr]^{-1},
\end{equation}
or equivalently,
\begin{eqnarray}\label{rochi3D}
\rho\tilde{\alpha}&\simeq&\frac{3\tilde{\chi}_{eff}}{\tilde{\chi}_{eff}+3}\Bigl[1-3(4\pi\xi^{3}\rho)^{-2}\tilde{\chi}_{eff}^{3}\nonumber\\&\times&
[9(4\pi\xi^{3}\rho)^{-2}\tilde{\chi}_{eff}^{3}+(\tilde{\chi}_{eff}+3)^{3}/3]^{-1}\Bigr].
\end{eqnarray}
Using the above relations, the modified $\tilde{\gamma}$-factors read,
\begin{widetext}
\begin{eqnarray}
2\tilde{\gamma}_{\perp}^{FF}(k)&\simeq&-i\frac{k}{2\pi}\frac{\tilde{\epsilon}_{eff}+2}{3}\Bigl[1-3(4\pi\xi^{3}\rho)^{-2}\tilde{\chi}_{eff}^{3}
[9(4\pi\xi^{3}\rho)^{-2}\tilde{\chi}_{eff}^{3}+(\tilde{\chi}_{eff}+3)^{3}/3]^{-1}\Bigr]^{-1}\:\sqrt{\tilde{\epsilon}_{eff}},\label{2gtff}\\
2\tilde{\gamma}_{\perp}^{NF}&=&\frac{-k}{2\pi\zeta}
\Bigl[\frac{1}{2}\tilde{\chi}_{eff}-\frac{3}{10}\tilde{\chi}_{eff}^{2}+[\frac{1}{2}(4\pi\xi^{3}\rho)^{-2}+0.181]\tilde{\chi}_{eff}^{3}
-[\frac{14}{15}(4\pi\xi^{3}\rho)^{-2}+0.11]\tilde{\chi}_{eff}^{4}...\Bigr],\label{g2mod}\\
\tilde{\gamma}_{\parallel}&\simeq&\frac{-k}{2\pi}\frac{1}{\zeta^{3}}
\Bigl[\tilde{\chi}_{eff}+\frac{1}{3}\tilde{\chi}_{eff}^{2}+[(4\pi\xi^{3}\rho)^{-2}-\frac{1}{8}]\tilde{\chi}_{eff}^{3}
+0.073\tilde{\chi}_{eff}^{4}...\Bigr]\label{g3mod}\\
&-&\frac{k}{2\pi}\frac{1}{\zeta}\Bigl[\frac{1}{2}\tilde{\chi}_{eff}+\frac{1}{2}\tilde{\chi}_{eff}^{2}+
[\frac{1}{2}(4\pi\xi^{3}\rho)^{-2}-0.076]\tilde{\chi}_{eff}^{3}...\Bigr]\label{g4mod}\\
&-&i\frac{k}{2\pi}\Bigl[\frac{1}{3}\tilde{\chi}_{eff}+\frac{1}{3}\tilde{\chi}_{eff}^{2}+
[\frac{1}{3}(4\pi\xi^{3}\rho)^{-2}-0.05]\tilde{\chi}_{eff}^{3}...\Bigr]\label{g5mod}.
\end{eqnarray}
\end{widetext}
There is just a subtle point about the $\mathcal{O}(\zeta^{-1})$ and $\mathcal{O}(\zeta^{0})$ terms of order $(\tilde{\alpha}/4\pi\xi^{3}\rho)^{2}\tilde{\chi}_{eff}^{3}$ in Eqs.(\ref{g2mod},\ref{g4mod},\ref{g5mod}). Differently to the expression for $\chi^{(2)}_{\parallel MG^{ov}}$ in Eq.(\ref{xi2paral}), Eq.(\ref{xiparalrec}) does not contain terms of orders $\mathcal{O}(\zeta^{2},\zeta^{3})$. Therefore, the substitution of $\tilde{\chi}_{\parallel}^{(2)}(Q)$ instead of $\chi_{\parallel MG^{ov}}^{(2)}(Q)$ in Eq.(\ref{laGpar}) does not yield directly all the terms of orders $\mathcal{O}(\zeta^{-1},\zeta^{0})$ in $\tilde{\gamma}_{\parallel}$. The reason being that in the formulae for $\bar{\tilde{\chi}}^{(2)rec.}_{st.,sc.}$ we have disregarded the coupling between electrostatic and radiative bare propagators. However, exploiting the similarity pointed out after Eq.(\ref{laqtoca}) between, in this case, $\tilde{\gamma}_{\parallel}$ and  $2\tilde{\gamma}_{\perp}^{NF}$, those terms can be derived. The coincidence between the longitudinal and transverse terms at order  $\rho\tilde{\alpha}(\tilde{\alpha}/4\pi\xi^{3})^{2}$ (equivalently, at order $(\tilde{\alpha}/4\pi\xi^{3}\rho)^{2}\tilde{\chi}_{eff}^{3}$) is depicted diagrammatically in Fig.\ref{MG4}.\\
\indent It is worth noting that, for a given order in $\tilde{\alpha}$ or $\tilde{\chi}_{eff}$, those terms which include recurrent scattering factors $\tilde{\alpha}/4\pi\xi^{3}$ may be greater than those involving just $\rho\tilde{\alpha}$ factors. The weight of $\tilde{\alpha}/4\pi\xi^{3}$ relative to $\rho\tilde{\alpha}$ can be computed as a function of the effective filling factor, $\phi$. Treating the scatterers as hard spheres of radius $\xi/2$, regardless of the internal radius $a$ of the scatterers, $\rho=\frac{6\phi}{\pi\xi^{3}}$. Thus we find $4\pi\xi^{3}\rho=24\phi$, which is less than one for $\phi\lesssim0.05$.

\subsection{Correction to the MG coherent emission}

Recurrent scattering introduces also a modification on the local field factor \emph{w.r.t.} that of Lorentz-Lorenz. Correspondingly, the expressions for LDOS$^{Pols.}_{\perp}(\omega)$ and $W^{Coh.}(\vec{r})$ get also modified \emph{w.r.t.} Eqs.(\ref{LDOSMG},\ref{wcohMG}). Using the relation of Eq.(\ref{rochi3D}), they read now,
\begin{widetext}
\begin{eqnarray}
\tilde{\textrm{LDOS}}^{Pols.}_{MG\perp}(\omega)&=&\frac{\omega^{2}}{4\pi^{2} c^{2}}\:\Re{\Bigl\{\frac{\tilde{\epsilon}_{eff}+2}{3}\Bigl[1-3(4\pi\xi^{3}\rho)^{-2}\tilde{\chi}_{eff}^{3}
[9(4\pi\xi^{3}\rho)^{-2}\tilde{\chi}_{eff}^{3}+(\tilde{\chi}_{eff}+3)^{3}/3]^{-1}\Bigr]^{-1}\Bigr\}}\Re{\{\sqrt{\tilde{\epsilon}_{eff}}\}},\label{LDOSREC}\\
\tilde{W}^{Coh.}(\vec{r})&\simeq&\mathcal{W}^{0}_{\omega}\:\Bigl|3\frac{\tilde{\epsilon}_{eff}-1}{\tilde{\epsilon}_{eff}+2}\Bigr|^{2}\:\Bigl|1-3(4\pi\xi^{3}\rho)^{-2}\tilde{\chi}_{eff}^{3}
[9(4\pi\xi^{3}\rho)^{-2}\tilde{\chi}_{eff}^{3}+(\tilde{\chi}_{eff}+3)^{3}/3]^{-1}\Bigr|^{2}\nonumber\\
&\times&\Re{}\Bigl\{\frac{\tilde{\epsilon}_{eff}+2}{3}\Bigl[1-3(4\pi\xi^{3}\rho)^{-2}\tilde{\chi}_{eff}^{3}
[9(4\pi\xi^{3}\rho)^{-2}\tilde{\chi}_{eff}^{3}+(\tilde{\chi}_{eff}+3)^{3}/3]^{-1}\Bigr]^{-1}\Bigr
\}\Re{\{\sqrt{\tilde{\epsilon}_{eff}}\}},\label{wcohMOD}
\end{eqnarray}
\end{widetext}

\section{Experimental proposals}\label{Experiments}
In principle, numerical simulations based on the Coupled-Dipole-Method (CDM) \cite{Purcell,Draine} or the Boundary-Element-Method (BEM) \cite{GarciadAbajo} could test our analytical results. However, to the author's knowledge, there is no numerical code that takes into account consistently all the radiative corrections on $\tilde{\alpha}$ and $\chi_{eff}$. Some numerical works consider only electrostatic corrections due to actual recurrent scattering \cite{Mukamel,Feldoher3}. Others do include radiative corrections but only those in free-space \cite{Sentenac}. Some authors focus on deviations \emph{w.r.t.} the Claussius-Mossotti formula due to the finite size of the dipole constituents \cite{Latakhia}. An improvement was made by the authors of  Ref.~\onlinecite{Draine2} who carried out an a priori renormalization of the polarizability of point dipoles in a lattice by including the lattice dispersion relation. Further on, the authors of Ref.~\onlinecite{Sentenac3} have proposed a modification over the original CDM formulation which accounts for the integration in $\tilde{\alpha}$ of the near-field corrections within a lattice cell. Other works include additional radiative corrections on $\chi_{eff}$. In Ref.~\onlinecite{Feldoherp} retardation effects for $\zeta\sim1$  are considered. In Ref.~\onlinecite{Sentenac2} the corrections come from the finiteness of the macroscopic sample.\\
\indent In the following, we propose both numerical and laboratory experiments to test our results. In the first place, we intend to verify the accuracy of the self-consistent overlap MG and recurrent approximations in the computation of the radiative corrections on a sample made of hard-sphere scatterers. Second, we aim to verify the generic form of our formulae for $W^{Coh.}_{MG}(\vec{r})$ and $\tilde{W}^{Coh.}(\vec{r})$ in an MG dielectric. The latter formulae are the  most universal result of this article as they are nearly independent of the microscopical details of the medium. That is, they are the result of considering only the inherent correlations --i.e., exclusion volume and self-correlation-- whose detailed form is almost irrelevant for long wavelength propagating modes.
\subsection{Test of the approximations}
\subsubsection{Numerical experiments on emission}
Despite the fact that numerical simulations do not take proper account of all the radiative corrections, the formula derived in  Ref.~\onlinecite{PRAvc} for the total power emitted/absorbed by an induced dipole is still testable by means of CDM numerical experiments. It was found in Ref.~\onlinecite{PRAvc} that
\begin{eqnarray}\label{laWshort}
W^{Tot}_{\omega}&=&
\frac{-\omega^{3}\epsilon_{0}}{6c^{2}}|\tilde{\alpha}E_{0}^{\omega}|^{2}\Bigl[\Im{\{2\gamma^{(0)}_{\perp}+2\gamma_{\perp}
+\gamma_{\parallel}\}}\\&-&\frac{3}{k^{2}}\frac{\Im{\{\alpha_{0}\}}}{|\alpha_{0}|^{2}}\Bigr].\nonumber
\end{eqnarray}
If radiative corrections were properly incorporated in the renormalized polarizability, the above equation would be equal to $\frac{\omega\epsilon_{0}}{2}|E_{0}^{\omega}|^{2}\Im{\{\tilde{\alpha}\}}$. Such an equivalence is a consequence of the optical theorem in a complex medium. Nonetheless, even if radiative corrections are not consistently incorporated in $\tilde{\alpha}$, Eq.(\ref{laWshort}) can be tested numerically in aggregates of point dipoles like those of Mallet, Gu\'{e}rin and Sentenac \cite{Sentenac}. The reason is as follows. In Eq.(\ref{g1}) and Eq.(\ref{2gtff}) the $\gamma$-factors can be written in function of $\chi_{MG}$ and $\tilde{\chi}_{eff}$ because $\chi_{MG}$ and $\tilde{\chi}_{eff}$ are related to $\tilde{\alpha}$ by Eq.(\ref{MGeq}) and Eq.(\ref{rochi3D}) respectively. Likewise, we can use Eqs.(\ref{MGeq},\ref{rochi3D}) to replace  $\tilde{\alpha}E_{0}^{\omega}$ in Eq.(\ref{laWshort}) with a function of $\chi_{MG},\vec{E}^{\omega}_{0}(\vec{r})$ or $\tilde{\chi}_{eff},\vec{E}^{\omega}_{0}(\vec{r})$ respectively. That is irrespective of the manner $\tilde{\alpha}$ is renormalized.  In particular, $\tilde{\alpha}$ can be taken to be the in-free-space value, $\alpha$, as used in some simulations \cite{Sentenac,LuisRemiMole}. By exciting one of the dipoles in the aggregate with an external probe field as in Ref.~\onlinecite{LuisRemiMole,LuisFroufe2}, integration of the total intensity in the far field must adjust to the first term of Eq.(\ref{laWshort}), with $\tilde{\alpha}$ replaced with $\alpha$ and disregarding microscopical absorbtion in the medium. The latter means that the macroscopic absorbtion is just due to microscopic incoherent scattering in agreement with the interpretation of the authors of  Ref.~\onlinecite{Sentenac}. As mentioned above, the $\zeta$-dependent terms are only exact at $\mathcal{O}(1)$ in $\chi_{MG},\tilde{\chi}_{eff}$. Therefore, this would test the accuracy of the self-consistent overlap MG and recurrent approximations in the near field.
\subsubsection{Laboratory experiments on transmission}\label{bulkExp}
Direct verification of our approach needs of laboratory experiments. Accurate information of the microscopic parameters $\alpha_{0}$, $k_{0}$ and $\xi$ is required as well as isolation of the radiative effects from those which are non-radiative.\\
\indent Experimental measurements are performed on the propagation  of a normally incident light beam through a dielectric slab. The refractive index, $n$, and the extinction coefficient, $\kappa$, are derived out of the formulae of the transmition/reflection coefficients. Because the effects of radiative corrections are expected to be more relevant under quasi-resonant conditions, instead of absolute measurements of $n$ and $\kappa$ experiments concentrate on their spectra close to the resonance. The signatures of the radiative effects under investigation are the resonant frequency shift and the band broadening \emph{w.r.t.} the values of the isolated dipoles in free space. According to our formulation, the frequency shift must be given by Eq.(\ref{kres}) plus the Lorentz-shift, $-\frac{1}{3}\rho\alpha_{0}k_{0}^{2}$, in the MG model. The latter must be corrected by an additional term $-\frac{1}{3}\rho\alpha_{0}k_{0}^{2}(\frac{\tilde{\alpha}}{4\pi\xi^{3}})^{2}$ when recurrent scattering is considered. The line-width must adjust to Eq.(\ref{Gamares}).\\
\indent A pioneering experiment on this issue is that of Ref.~\onlinecite{PRLMakietal} on an atomic gas of potassium. There, selective-reflection techniques were employed to measure the frequency shift of the transition  $4\:^{2}\textrm{S}_{1/2}\rightleftarrows4\:^{2}\textrm{P}_{1/2}$. The problem to test our formalism in that setup is that both the Doppler broadening and other collisional effects dominate over the radiative effects. A good estimate for  $\xi$ is also difficult to obtain. For cold atoms it would be of the order of twice the van-der-Waals radius. At high temperature however it is affected by the collision cross-section of the atoms.\\
\indent On the opposite extreme is an atomic glass. It has the advantage that $\xi$ can be accurately determined and vibrational effects well controlled. However, atomic orbitals overlap in such a way that the electronic band configurations can be very different to those of the individual atoms isolated. That makes the estimate of $\alpha_{0}$ difficult to achieve as its value may be very different to that of isolated dipoles. On top of that, there might be contributions of free electrons, which introduce an additional source of uncertainty in computing the component of the susceptibility due to plasma oscillations \cite{Potter}.\\
\indent We conclude that suitable experimental setups must be such that undesired collisional effects reduce to a minimum; control must be kept over the parameter $\xi$; and the electronic structure of the dipole-constituents must remain unaltered either by their mutual cohesive interactions or by their interaction with a substrate. All these conditions meet in 'artificial solids' like those prepared by Liz-Marz\'{a}n, Giersig, Mulvaney and Ung \cite{Liz1,Liz0,Liz2}. There, the dipole constituents are gold nanoparticles. The particles are first coated by silica shells prior to their layer-by-layer assembly. That way, the nanoparticles do not form aggregates which could modify their single-particle electronic structure otherwise. Because the metal is nobel, no chemical reaction exists with the substrate. As a result, the dielectric properties of the isolated gold particles are well described by Drude's model prior to formation of the solid and a good estimation for $\xi$ can be obtained. In Ref.~\onlinecite{Liz2} the authors carried out experimental measurements of the absorbance and the plasmon resonant shift in thin films as a function of the filling factor. The exclusion volume is well approximated by twice the silica shell-thickness plus the radius of the Au particles. The density and size of the particles are known by construction. The results are compared with the ordinary MG formulae. The agreement is qualitative but not quantitative enough. Since our approach incorporates the near field terms which are missing in the ordinary MG formula and the coated particles are perfectly described as hard-spheres, we expect to obtain a better fit. However, because the present work covers only the off-resonance part of the spectrum of $n(\omega)$ and $\kappa(\omega)$, the at-resonance curves of  Ref.~\onlinecite{Liz2} cannot be compared with our formulae. Nevertheless, a similar experimental setup might still be useful to verify our formulae for the spectrum of coherent emission as explained below.
\subsection{Test of the Coherent Emission Spectrum}\label{Cohitest}
For practical purposes, the measurement of the coherent intensity is of interest in signal processing and transmission at both macroscopical and microscopical scales \cite{nanoantenas}. Also, there is a fundamental reason why the verification of our result is of interest. That is, the formulae of $W^{Coh.}_{MG}(\vec{r})$ and $\tilde{W}^{Coh.}(\vec{r})$ are nearly independent of the microscopical details of the medium for low frequency emission. Therefore, they can be considered as generic for the kind of media which adjust to the so-called \emph{virtual cavity scenario}. Those are, media made of a collection of equivalent dipoles in which the emitter is itself one of the host dipoles. The verification of our result would throw light into the historical problem of the role of the local field factors in dipole emission.
\subsubsection{Numerical experiments}
The formula of Eq.(\ref{wcohMG}) for $W^{Coh.}_{MG}(\vec{r})$ is exact for the s.s. MG model since the overlap approximation is exact in the long-wavelength limit, $q\xi\rightarrow0$ \cite{Feldoher,vanTigg}. When corrections due to the inherent recurrent scattering are incorporated, the formula in Eq.(\ref{wcohMOD}) for $\tilde{W}^{Coh.}(\vec{r})$ is exact up to order $\mathcal{O}(\tilde{\chi}_{eff}^{5})$ for a hard-sphere model. Because both $W^{Coh.}_{MG}(\vec{r})$ and $\tilde{W}^{Coh.}(\vec{r})$ depend only on macroscopic quantities, Eqs.(\ref{wcohMG},\ref{wcohMOD}) do not need of the knowledge of any microscopical parameter to be validated\footnote{Except for the dependence of $\tilde{W}^{Coh.}(\vec{r})$ on $\xi$ which can be replaced by its relation with $\phi$.}. In numerical simulations it is obvious that the access to microscopical parameters is not a problem. On the contrary, in experimental setups it is, and that is why the validation of Eqs.(\ref{wcohMG},\ref{wcohMOD}) would be easier than the direct validation of $\epsilon_{MG}$ and $\tilde{\epsilon}_{eff}$. We sketch first  how to verify Eqs.(\ref{wcohMG},\ref{wcohMOD}) in the numerical simulations with dipole aggregates of the kind of those in Ref.~\onlinecite{Sentenac,LuisRemiMole}. As for $W^{^{Tot}_{\omega}}$, the dipole at $\vec{r}$ must be excited by an external field. Then,
it is necessary to compute statistically the \emph{average} coherent field. Performing the statistical average at a single point should be enough provided that the orientation of the emitter is randomized as well \cite{LuisFroufe2}. Differently to the measurement of the total radiation which can be performed out of a finite size sample, the point $\vec{r}'$ where the coherent field is measured must lie well inside the sample and in the far field with $|\vec{r'}-\vec{r}|\gg\xi$. Randomization of the emitter orientation implies that the intensity associated to the average field at $\vec{r}'$ is the angular-average intensity at a distance $|\vec{r'}-\vec{r}|$ from the emitter. Thus, the square of that field times $2\pi |\vec{r'}-\vec{r}|^{2}c\epsilon_{0}n$ is the quantity to compare with the Beer-Lambert formula of Eq.(\ref{WCohprop}).
\subsubsection{Laboratory experiments}\label{Tipsec}
\begin{figure}[h]
\includegraphics[height=5.0cm,width=7.0cm,clip]{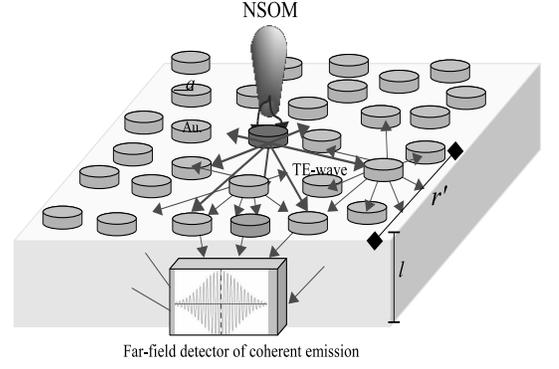}
\caption{Schematic representation of the experimental setup suggested in Section \ref{Tipsec}. The p-polarized near-field of an NOMS is used to excite one of the gold nanorods of radius $a$ and length $l$, with $k^{-1}\gg l\gg a$ . The on-plane polarized radiation (TE-modes) experiences multiple-scattering processes until collected in the far-field. The detector is placed right at he edge of the sample at a distance $r'$ away from the emitter. The coherent signal is filtered using  interferometry.}\label{Tip}
\end{figure}

The preparation of a suitable laboratory experiment to test Eqs.(\ref{wcohMG},\ref{wcohMOD}) involves a number of difficulties. In principle, independent measurements of the refractive index, the extinction coefficient and $W^{Coh.}(\vec{r})$ could be performed. To determine $n$, $\kappa$, standard transmittion/reflection measurements on monochromatic light can be carried out. For the computation of $W^{Coh.}(\vec{r})$, a unique dipole must be excited and the far field coherent radiation of Eq.(\ref{WCohprop}) must be detected. For the latter, the detector must be introduced in a manner that it does not modify the surrounding dielectric medium. For instance, it can be placed right at the edge of the sample. Coherent radiation can be filtered using interferometry techniques. Unfortunately, there exists a potential difficulty in the excitation of the emitter. Should the emitter be an isolated particle in free space or a single particle doped with a fluorescent ion, a laser beam could be used as in Ref.~\onlinecite{Shniepp}. However, the stimulation of a unique dipole in an MG dielectric using an incident light beam is not possible due to the inherent diffraction limit. That is, a monochromatic light beam cannot resolve distances less than $\sim\lambda/2$. Therefore, it would be necessary that the radius of the exclusion volume, $\xi$, be at least of the order of $\lambda$ to avoid the excitation of more than one dipole. However, the low frequency requirement implies $\xi/\lambda\ll1$ and so both conditions cannot be satisfied at the same time. As an alternative, we propose the use of near-field optical techniques \cite{NOMS}. That implies the introduction of an external device inside the dielectric, namely the metallic tip  of a near-field-scanning-optical-microscope (NOMS). The tip excites the emitter dipole via near-field induction. In a three-dimensional dielectric that would break manifestly translation invariance since the tip would act as an impurity. That is, the tip could scatter part of the emitted radiation and also interact with the dipoles around the emitter. Thus, unpredictable deviations from Eqs.(\ref{wcohMG},\ref{wcohMOD}) are expected. To go around this problem, we rather propose to work on a quasi-two-dimensional dielectric. The tip must be placed perpendicularly above the sample as in Fig.\ref{Tip} so that it does not perturb the medium. In order to avoid the interaction of the emitted radiation with the tip, only on-plane polarized modes  --i.e., p-polarization or TE-modes-- are of our interest. An analogous formula to that in Eq.(\ref{wcohMG}) is worked out in Appendix \ref{Append} for the TE coherent emission per unit length, $l$,
\begin{eqnarray}\label{wcohMG2D}
l^{-1}W^{Coh.}_{pMG}(\vec{r})&=&\mathcal{W}^{p0}_{\omega}\:\Bigl|2\frac{\epsilon^{p}_{MG}-1}{\epsilon^{p}_{MG}+1}\Bigr|^{2}\\
&\times&\frac{\Re{\{\epsilon^{p}_{MG}\}}+1}{2}\Bigl[1+\pi^{-1}\arctan{\Bigl(\frac{\Im{\{\epsilon^{p}_{MG}\}}}{\Re{\{\epsilon^{p}_{MG}\}}}\Bigr)}\Bigr],\nonumber
\end{eqnarray}
where the script $p$ stands for p-polarization and  $\mathcal{W}_{\omega}^{p0}=\frac{\epsilon_{0}\omega^{3}}{16c^{2}\sigma^{2}}|\vec{E}^{\omega}_{0}(\vec{r})|^{2}$ with $\sigma$ being the numerical surface density of dipoles. The preparation of two dimensional samples could be achieved using the same coating techniques as for three dimensional artificial solids \cite{Rods}. Alternatively, regular arrays of the kind of those in  Ref.~\onlinecite{Soukoulis} can be used.\\
\indent Beyond the MG approximation a formula for $l^{-1}\tilde{W}^{Coh.}_{p}(\vec{r})$ can also be found which captures the effects of recurrent scattering up to $\mathcal{O}[(\tilde{\chi}^{p}_{eff})^{5}]$. We proceed in the same manner as for $\tilde{W}^{Coh.}(\vec{r})$ in Eq.(\ref{wcohMOD}). The two-dimensional expressions analogous to Eqs.(\ref{tila},\ref{rochi3D}) are respectively,
\begin{equation}
\tilde{\chi}^{p}_{eff}=\rho\tilde{\alpha}\Bigl[1-\frac{1}{2}\sigma\tilde{\alpha}_{p}[1+(\tilde{\alpha_{p}}/2\pi\xi^{2})^{2}]\Bigr]^{-1},
\end{equation}
\begin{eqnarray}\label{rochi2D}
\sigma\tilde{\alpha}_{p}&\simeq&\frac{2\tilde{\chi}^{p}_{eff}}{\tilde{\chi}^{p}_{eff}+2}\Bigl[1-2(2\pi\xi^{2}\sigma)^{-2}(\tilde{\chi}^{p}_{eff})^{3}\nonumber\\
&\times&[6(2\pi\xi^{2}\sigma)^{-2}(\tilde{\chi}^{p}_{eff})^{3}+(\tilde{\chi}^{p}_{eff}+2)^{3}/2]^{-1}\Bigr].
\end{eqnarray}
And so, by replacing the local field factor with its recurrent-corrected value we end up with,
\begin{widetext}
\begin{eqnarray}\label{wcohMod2D}
l^{-1}\tilde{W}^{Coh.}_{p}(\vec{r})&=&\mathcal{W}^{p0}_{\omega}\:\Bigl|2\frac{\tilde{\epsilon}^{p}_{eff}-1}{\tilde{\epsilon}^{p}_{eff}+1}\Bigr|^{2}
\:\Bigl|1-2(2\pi\xi^{2}\sigma)^{-2}(\tilde{\chi}^{p}_{eff})^{3}
[6(2\pi\xi^{2}\sigma)^{-2}(\tilde{\chi}^{p}_{eff})^{3}+(\tilde{\chi}^{p}_{eff}+2)^{3}/2]^{-1}\Bigr|^{2}\nonumber\\
&\times&\Re{}\Bigl\{\frac{\tilde{\epsilon}^{p}_{eff}+1}{2}\Bigl[1-2(2\pi\xi^{2}\sigma)^{-2}(\tilde{\chi}^{p}_{eff})^{3}\:
[6(2\pi\xi^{2}\sigma)^{-2}(\tilde{\chi}^{p}_{eff})^{3}+(\tilde{\chi}^{p}_{eff}+2)^{3}/2]^{-1}\Bigr]^{-1}\Bigr\}\nonumber\\
&\times&\Bigl[1+\pi^{-1}\arctan{\Bigl(\frac{\Im{\{\tilde{\epsilon}^{p}_{eff}\}}}{\Re{\{\tilde{\epsilon}^{p}_{eff}\}}}\Bigr)}\Bigr].
\end{eqnarray}
\end{widetext}
The verification of Eqs.(\ref{wcohMG2D},\ref{wcohMod2D}) in numerical simulations is also computationally less demanding than those in three dimensions --see eg. Ref.~\onlinecite{Silvia,Vynck}.
\subsection{A numerical example}\label{ex}
\begin{figure}[h]
\includegraphics[height=10.0cm,width=8.6cm,clip]{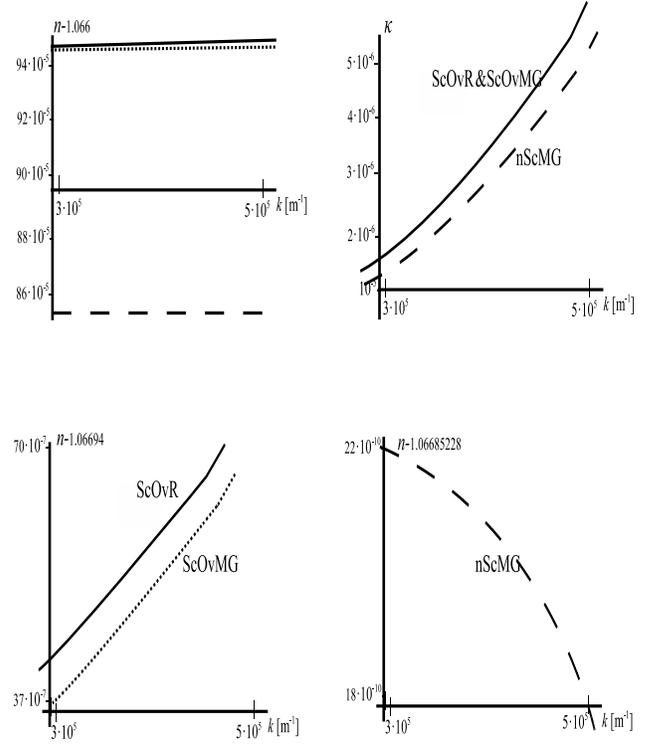}
\caption{Refractive index, $n$, and extinction coefficient, $\kappa$, for the numerical example of Section \ref{ex}. The long-dashed line represents the non-self-consistent MG approximation (nScMG). The short-dashed line corresponds to the self-consistent overlap MG approximation (ScOvMG). The solid line represents the self-consistent overlap recurrent approximation (ScOvR) of Section \ref{beyond}. The curves of $n$ appear separated in the lower plots.}\label{MG5}
\end{figure}

\begin{figure}[h]
\includegraphics[height=5.0cm,width=8.6cm,clip]{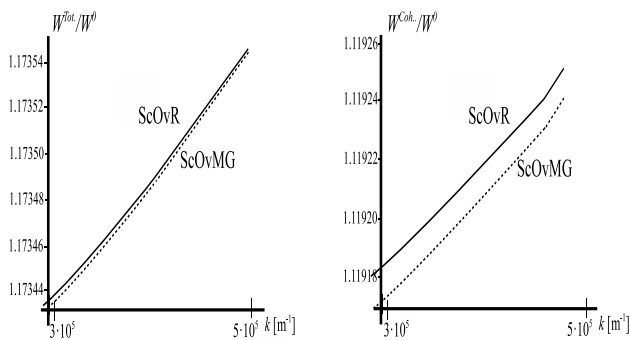}
\caption{Plots of the total and coherent stimulated emission in the self-consistent overlap MG approximation (ScOvMG, short-dashed line) and in the self-consistent overlap recurrent approximation (ScOvR, solid line). Quantities are normalized to the in-free-space emission, $W^{0}$.}\label{MG6}
\end{figure}

We perform here the numerical computation of the dielectric constant of an MG dielectric made of classical hard-sphere dipoles. Also, we compute the power emitted by one of the dipoles when stimulated by an external stationary monochromatic field. The dipoles are spheres of radius $a=0.1\mu m$ and real dielectric constant $\epsilon_{e}=16$. For this model, the bare electrostatic polarizability is $\alpha_{0}=4\pi\frac{\epsilon_{e}-1}{\epsilon_{e}-2}a^{3}\simeq1.11\cdot10^{-20}m^{3}$. The wavelengths of the probe field range from $300nm$ to $500nm$ which correspond to frequencies between $3\cdot10^{5}m^{-1}c$ and $5\cdot10^{5}m^{-1}c$, much lower than any resonant frequency by assumption. To ensure that both the dipole approximation and the low frequency condition hold, the radius of exclusion volume is set to $\xi=0.4\mu m$. In order to avoid overdensity terms in the correlation function and, at the same time, allow for multiple-scattering, we set $\phi=0.4$. Note that because $\xi=4a$, the material filling fraction is $8$ times less, $0.05$. This corresponds to a numerical density $\rho=1.49\cdot10^{19}m^{-3}$. We perform the computation of the refractive index and the extinction coefficient in three approximations. In the first one, a non-self-consistent MG approximation is used. In it, single particle polarizabilities are only affected by in-free-space radiative corrections,  $\alpha=\alpha_{0}[1-\frac{i}{6\pi}k^{3}\alpha_{0}]^{-1}$, and the effective susceptibility adjusts to that of the s.s. MG model. In the second one, the self-consistent overlap MG approximation is employed and the $\gamma$-factors computed in Section \ref{lasgamaov} are used to renormalize both the single-particle polarizability and the dielectric constant. In the third one, the self-consistent overlap recurrent approximation of Section \ref{beyond} is used. The results are plotted in the graphs of Fig.\ref{MG5}.\\
\indent We have also computed the total power emitted by a random dipole when stimulated by a stationary monochromatic field in the same range of frequencies. Eq.(\ref{laWshort}) is applied for $W^{Tot.}$. For the computation of the coherent emission, $W^{Coh.}$, we use Eqs.(\ref{wcohMG},\ref{wcohMOD}). The results are plotted in the graphs of Fig.\ref{MG6}. The values are normalized to that of the in-free-space emission, $W^{0}=\frac{\omega}{2}\epsilon_{0}\Im{\{\alpha\}}$.
\section{Discussion}\label{discus}
\subsection{Corrections at high filling fractions}\label{limit}
When the effective filling fraction of scatterers is relatively high, the first-neighbor overdensity term in  $h(r)$  becomes relevant for the computation of $\bar{\chi}^{(2)}$.  That is the case when the relation $\xi/\rho^{-1/3}$ approaches one. The functional form of the overdensity  term  depends on the precise profile of the interaction potential between dipoles. For the sake of simplicity we will restrict to the delta function of Eq.(\ref{darriba}), $\Delta h(r)=C\:\xi\delta^{(1)}(r-\xi)$. The inclusion of such a term in the computation of $\bar{\chi}^{(2)}(Q)$ yields the additional contributions,
\begin{equation}
\Delta\chi^{(2)}_{\perp}(Q)\simeq-C(\rho\tilde{\alpha})^{2}[\frac{2}{3}\zeta^{2}+\frac{1}{15}Q^{2}+\mathcal{O}(\zeta^{3},\zeta^{2}Q^{2})+..],
\label{Dxiperp}
\end{equation}
\begin{equation}
\Delta\chi^{(2)}_{\parallel}(Q)\simeq-C(\rho\tilde{\alpha})^{2}[\frac{2}{3}\zeta^{2}-\frac{2}{15}Q^{2}+\mathcal{O}(\zeta^{3},\zeta^{2}Q^{2})+..].
\label{Dxiparal}
\end{equation}
The relevance of these terms in our previous formulae is as follows.
Regarding far-field calculations and coherent emission terms,
only long wavelength modes with $Q\ll1$  enter Eqs.(\ref{laGamaA},\ref{2gtff}). Those equations contain the order $\mathcal{O}(\zeta^{0})$ terms which amount to the transverse coherent modes. The inclusion of Eqs.(\ref{Dxiperp},\ref{Dxiparal}) is irrelevant for them. Therefore, the degree of accuracy of Eqs.(\ref{laGamaA},\ref{2gtff}) remains unaltered regardless of the degree of order in the medium. The same argument applies to Eqs.(\ref{wcohMG},\ref{wcohMOD}). On the contrary,  Eqs.(\ref{laGamaB},\ref{laqtoca}) and Eqs.(\ref{g2mod}-\ref{g5mod}) contain short wavelength modes with $Q\gtrsim1$. Therefore, in order to keep their degree of accuracy for high filling fractions, it is necessary the inclusion of additional terms. Those are, the ones of order $\mathcal{O}(\zeta^{0})$ of Eq.(\ref{Dxiperp}) in $\chi^{(2)}_{\perp MG^{ov}}, \tilde{\chi}^{(2)}_{\perp}$ and those of orders $\mathcal{O}(\zeta^{0})$, $\mathcal{O}(\zeta^{2})$ and $\mathcal{O}(\zeta^{3})$
of Eq.(\ref{Dxiparal}) in $\chi^{(2)}_{\parallel MG^{ov}}, \tilde{\chi}^{(2)}_{\parallel}$.\\
\indent Regarding experimental observations, in order to prevent high-order effects in the artificial solids,
 it would be necessary to deform the surface of the silica shells to avoid first-neighbor overdensities around each coated gold particle.
\subsection{Comparison with previous approaches}
It has been mentioned already  that numerical simulations do not take proper account of radiative corrections in the renormalization of the single-particle polarizability. Hence, the dipole emission cannot be evaluated by direct application of the optical theorem in complex media, Eq.(\ref{laWshort}), unless knowledge of $\chi_{eff}$ be implicitly assumed in some approximation --eg. Ref.~\onlinecite{Sentenac3}. Nonetheless, we have explained how both the total and coherent intensities can be evaluated in simulations by exciting one of the dipoles of a sample with an external probe field as in Ref.~\onlinecite{LuisRemiMole,LuisFroufe2}. Mallet \emph{et al.} \cite{Sentenac} have measured instead the coherent and incoherent intensity scattered by a finite sample. They have compared the numerical data with the corresponding formulae for a Mie sphere with an effective dielectric constant adjusted to the MG formula. Our reasoning in Section \ref{limit} would explain why the authors found good agreement for the coherent cross-section, even for high filling factors.
On the contrary, because the ordinary MG formula does not include the $\zeta$-dependent terms which enter the incoherent intensity (or the macroscopic absorption in the terminology of Ref.~\onlinecite{Sentenac}), it is not surprising the large discrepancy
found also there in the incoherent cross-section for high effective dielectric constant. Direct comparison of our predictions with the numerical data of Ref.~\onlinecite{Sentenac} is not possible as the authors deal with scattering by a Mie particle. It is however possible to verify that, in the single-scattering approximation and in the small particle limit, our decomposition between coherent and incoherent emission agrees with the calculation of Froufe \emph{et al.} \cite{LuisRemiMole}.\\
\indent Cichocki and Felderhof developed in Ref.~\onlinecite{Feldoher2} a stochastic approach to the dielectric constant which is formally very similar to ours. Our \emph{self-polarization} $\gamma$-factors of Eqs.(\ref{LDOSIparalg},\ref{LDOSIperg}) are analogous --but not physically equivalent-- to their \emph{reaction field operator}. Our two-scatterer susceptibility in the quasicrystalline approximation, $\bar{\chi}^{(2)}$, is an approximation to their \emph{short-range connector}. As a result, they obtained a formula for the dielectric constant which is of the same functional form as that in Eq.(\ref{laeps}) once the limit $q\rightarrow0$ is taken. A similar strategy was followed by Leegwater and Mukamel in Ref.~\onlinecite{Mukamel}. That is, they first dressed up the single particle polarizability including near-field interactions and then applied the MG formula over the renormalized polarizability.  Cichocki and Felderhof applied their formalism to the particular case of the Drude-Lorentz model (equivalently, the Maxwell-Garnett model) in a non-polar fluid \cite{ElectroFeld,Feldoher3,Feldoher4}. In the two latter references they concentrated on the resonant regime. For the computation of the reaction field operator they developed in Ref.~\onlinecite{Feldoher3} a self-consistent ring approximation which is similar to our self-consistent overlap recurrent approximation. However, our approximation is performed off resonance and for the sake of simplicity we restrict to $\mathcal{O}[\tilde{\alpha}^{2}/(4\pi\xi^{3})^{2}]$ terms in $\bar{\tilde{\chi}}^{(2)rec.}_{st.}$. Our approach includes transverse modes while the authors of Ref.~\onlinecite{ElectroFeld,Feldoher3,Feldoher4} restricted to electrostatic induction only. More importantly, because transverse modes are included neither in Ref.~\onlinecite{ElectroFeld,Feldoher3,Feldoher4} nor in Ref.~\onlinecite{Mukamel}, the radiative modes of Eqs.(\ref{laGamaA},\ref{2gtff}) are missing there and the computation of the radiative line-width cannot be complete. As a matter of fact, neither a proper renormalization of the single-particle polarizability that satisfies the optical theorem nor the formula for  $\tilde{W}^{Coh.}$ can be obtained following those approaches. Also, while the zero-modes which enter our $\tilde{\chi}_{eff}$ are properly accounted
for in Ref.~\onlinecite{ElectroFeld,Feldoher3,Feldoher4} since they are mainly affected by electrostatic interactions, only the $\mathcal{O}(\zeta^{-3})$ terms in our $\tilde{\gamma}_{\parallel}$
can be obtained out of those works.\\
\indent There are two fundamental differences between our approach and that of
Cichocki and Felderhof. The first one is technical. Because we have obtained in Ref.~\onlinecite{PRAvc} exact formulae for $\bar{\mathcal{G}}(\vec{r},\vec{r})$ as a function of $\bar{\chi}(q)$ --Eqs.(\ref{LDOSIper},\ref{LDOSIparal}), our strategy consists of computing first $\bar{\chi}(q)$ in some sensitive approximation. While its zero-mode is by definition $\chi_{eff}$, application of the exact formula for $\bar{\mathcal{G}}(\vec{r},\vec{r})$ yields the radiative corrections which enter  $\tilde{\alpha}$ . On the contrary, Cichocki and Felderhof developed in Ref.~\onlinecite{Feldoher3} an approximation to capture the most relevant recurrent scattering diagrams for the computation of $\tilde{\alpha}$. Those diagrams account for the multiple reflections that actual probe photons experience within clusters of particles, restricted to electrostatic induction. In their approximation, consecutive scattering processes are never uncorrelated and hence Dyson's propagator is never present. Because it is Dyson's propagator the relevant one for radiation, that approximation could not capture all the radiative corrections, but for those in free space at the most, even if radiative induction were implemented in their approach. In Ref.~\onlinecite{Feldoher4} the authors developed another approximation for the computation of their connector whose zero-mode enters $\chi_{eff}$. To this respect, ours is a lower order approximation.\\
\indent The second difference is conceptual and more subtle. The self-consistency requirement that the ring  approximations of
  Ref.~\onlinecite{Feldoher3,Feldoher4} satisfy are in perfect agreement with the application of the Couple-Dipole-Method. In fact, self-consistency is inherent
to the CDM equations. However, the self-consistency that we demand on radiative corrections
 apply to virtual photons instead. Hence, our comment at the beginning of Section \ref{Experiments}. That is, it is recognized that the application of the CDM equations
 needs of some a priori prescription for the value of the radiatively-corrected single-particle polarizability in order to obtain physically acceptable values for the scattering cross-section --see eg. Ref.~\onlinecite{Sentenac3}.
Our approach reveals that self-consistency in radiative corrections requires to solve for $\tilde{\alpha}$
 at the same time that the CDM system of equations is solved \cite{PRAvc}.\\
\indent Nevertheless, the elegant analytical approach developed in Ref.~\onlinecite{Feldoher3,Feldoher4}
will pave the way for the computation of the radiative corrections at resonance in a future publication.
\subsection{Relationship between the emission spectrum and the local field factors}\label{lff}
\begin{figure}[h]
\includegraphics[height=5.0cm,width=8.6cm,clip]{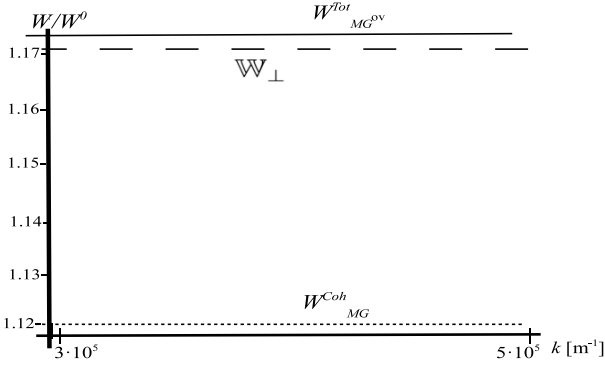}
\caption{Plots of the total and coherent stimulated emission for the numerical example of Section \ref{ex}. The solid line corresponds to the total emission in the overlap self-consistent MG approximation. The long-dashed line represents the transverse emission according to Eq.(\ref{wrong}). The short-dashed line corresponds to the coherent emission in the  overlap self-consistent MG approximation. Quantities are normalized to the in-free-space emission, $W^{0}$.}\label{MG7}
\end{figure}

In Section \ref{Cohitest} it has been emphasized the nearly universal character of Eqs.(\ref{wcohMG},\ref{wcohMG2D}) and Eqs.(\ref{wcohMOD},\ref{wcohMod2D}) in the low frequency limit of the virtual cavity scenario. Related with this fact, there exists a strong motivation to validate Eqs.(\ref{wcohMG},\ref{wcohMG2D}) and Eqs.(\ref{wcohMOD},\ref{wcohMod2D}). That is to show that the formulae for coherent emission in the \emph{stricto sensu} virtual cavity scenario contain only one local field factor,  $\Re{\{\frac{\chi_{\perp}(k_{\perp}^{Pol.})}{\rho\tilde{\alpha}}\}}$, and not two as commonly claimed in the literature --eg. Ref.~\onlinecite{LaudonPRL,LaudonJPB,KnollBarnett,Welsh,Juzel,PRLdeVries,Fleischhauer,Crenshaw,Schuurmans1}. In the following we will restrict ourselves to the s.s.  MG model. For this model the relations $\Re{\{\frac{\chi_{\perp}(k_{\perp}^{Pol.})}{\rho\tilde{\alpha}}\}}=\frac{\Re{\{\epsilon_{MG}\}}+2}{3}$, $\Re{\{\frac{\chi^{p}_{MG}}{\sigma\tilde{\alpha}_{p}}\}}=\frac{\Re{\{\epsilon^{p}_{MG}\}}+1}{2}$ are exact in three and two dimensions respectively and so are Eqs.(\ref{wcohMG},\ref{wcohMG2D}). Therefore, we have proved explicitly that only one Lorentz-Lorenz local field factor appears in the formula for coherent emission in an MG dielectric.\\
\indent In the literature, the usual expression in a three-dimensional dielectric for which an effective dielectric constant can be defined reads,
\begin{equation}\label{wrong}
\mathbb{W}_{\perp}\propto\Bigl(\frac{\epsilon_{eff}+2}{3}\Bigr)^{2}\sqrt{\epsilon_{eff}},
\end{equation}
where absorbtion is ignored.  The terminology varies from one author to another, referring to the above formula as virtual cavity \cite{Welsh}, Lorentz-Lorenz \cite{Fleischhauer} or Claussius-Mossotti \cite{KnollBarnett} emission\footnote{Most of those papers deal with the spontaneous decay rate instead. The distinction is irrelevant for the present discussion.}. Certainly none of those pioneer physicist worked out Eq.(\ref{wrong}) but just the relation between macroscopical classical fields in effective media \cite{Claussius,Mossotti,Lorentz,Lorenz}.\\
\indent It is clear that, if the above equation refers to the s.s. virtual cavity scenario and an effective medium exists, the medium must be an MG dielectric to which all the formalism developed throughout this work applies. In the papers cited above it is not clarified the coherent/incoherent nature of the radiation in Eq.(\ref{wrong}). However, the presence of the refractive index factor, $n=\sqrt{\epsilon_{eff}}$, denotes the integration of $\int\frac{\textrm{d}^{3}q}{(2\pi)^{3}}\:2\Im{\{G_{\perp}(q)\}}$ which would yield only coherent power. Thus, the expression in Eq.(\ref{wrong}) should stand for coherent emission. In any case, because the transverse incoherent emission in our calculation contains only terms of order $\mathcal{O}(\zeta^{-1})$, the addition of transverse coherent and transverse incoherent emission cannot equal Eq.(\ref{wrong}). Also contrary to the general assumption, effective longitudinal modes carry incoherent radiation even in the absence of absorption \cite{LaudonJPB}. That is proportional to the $\mathcal{O}(\zeta^{0})$ terms of $\gamma_{\parallel}^{MG^{ov}}$. The reason why some experiments on spontaneous emission seem to agree with Eq.(\ref{wrong}) \cite{Toptygin} is that those experiments measure total far-field radiation, and not only either coherent or transverse. It turns out that for $n$ close to one Eq.(\ref{wrong}) approximates the value of the total far-field radiation. That is the sum of Eq.(\ref{wcohMG}) plus longitudinal incoherent radiation. This is illustrated in Fig.\ref{MG7} where the data refer to the numerical example of Section \ref{ex}. Therefore, the rule out of Eq.(\ref{wrong}) in favor of our result could be verified by direct measurement of the coherent power. Note also that corrections due to recurrent scattering are of order $\mathcal{O}(\epsilon_{eff}^{3})$ and higher. Nevertheless, Eq.(\ref{wrong}) fails already at leading order.\\
\indent We stress that our conclusion about the presence of a unique local field factor in Eq.(\ref{wcohMG}) is not model-dependent. The equivalence $\Re{\{\frac{\chi_{\perp}(k_{\perp}^{Pol.})}{\rho\tilde{\alpha}}\}}=\frac{\Re{\{\epsilon_{MG}\}}+2}{3}$ is of course only valid for the s.s. MG model since it is implicit that both the condition $k\xi\ll1$ holds and the exclusion volume function is the only correlation function. Nonetheless, Eq.(\ref{WCOHA}) is exact in the s.s. virtual cavity scenario. That is, if the emitter can be treated as a single dipole which is in all equivalent to the rest of the constituents of the medium. As explained in Ref.~\onlinecite{PRAvc,Toptygin}, it is only for stimulated emission that the nature of the emitter remains equal to that of the rest of dipoles during the emission process. Hence, our proposal to measure $W^{Coh.}_{\perp}$ instead of $\Gamma^{Coh.}_{\perp}$.
\section{Conclusions}
We have computed the radiative corrections to the off-resonance dielectric constant of an MG dielectric. To that aim, we have used the formalism developed in Ref.~\onlinecite{PRAvc}. As a byproduct, we have obtained formulae for the spectrum of the coherent stimulated emission in both two and three dimensions. Two variants of the quasicrystalline approximation have been used. The first one is referred to as the self-consistent overlap  MG approximation. In it, the only relevant correlation between dipoles is approximated by a spherical exclusion volume. The effective susceptibility is that of the ordinary MG formula, Eq.(\ref{MGeq}). The $\gamma$-factors which enter the renormalization of $\tilde{\alpha}$ are those of Eqs.(\ref{g1}-\ref{g4}). The formulae for coherent radiation are those of Eq.(\ref{wcohMG}) and Eq.(\ref{wcohMG2D}) in three and two dimensions respectively. The second approximation is referred to as the  self-consistent overlap recurrent approximation. In it, in addition to the spherical exclusion volume, self-correlation is also considered and recurrent scattering diagrams are included up to $\mathcal{O}(\tilde{\alpha}^{2}/(4\pi\xi^{3})^2)$ in $\bar{\tilde{\chi}}^{(2)rec.}_{st.}$. The effective susceptibility is that of Eq.(\ref{tila}). The $\gamma$-factors which enter the renormalization of $\tilde{\alpha}$ are those of Eqs.(\ref{2gtff}-\ref{g5mod}). The formulae for coherent radiation are those of Eq.(\ref{wcohMOD}) and Eq.(\ref{wcohMod2D}) in three and two dimensions respectively. In Section \ref{Experiments} we have proposed several numerical and laboratory experiments to test our results. Most of them are based on the measurement of the coherent stimulated emission since the functional form of its formulae is nearly independent of the microscopical details of the medium. To this respect, we have found that only one local field factor enters those formulae in contradiction with the common assumption that two factors do.
\acknowledgments
We thank S. Albaladejo and L. Froufe for fruitful discussions, S. Martinez for proofreading the manuscript and B.U. Felderhof for
bringing to our attention some of the references. This work has been supported by the EU project
 NanoMagMa EU FP7-NMP-2007-SMALL-1 and the Scholarship Program 'Ciencias de la Naturaleza' of the Ramon Areces Foundation.
\appendix
\section{Coherent emission of TE modes in two dimensions}\label{Append}
In this section we derive Eq.(\ref{wcohMG2D}) for a two-dimensional dielectric made of cylindrical inclusions homogeneously distributed as in Fig.\ref{Tip}. Cylinders of radius $a\ll\lambda$ and permitivity $\epsilon_{e}$ have a bare classical and isotropic polarizability $\alpha^{0}_{p}=2\:\pi a^{2}\frac{\epsilon_{e}-1}{\epsilon_{e}+1}$ for p-polarized fields \cite{Silveirinha,Silvia}. The Maxwell-Garnett formula reads,
\begin{equation}\label{MG2D}
\chi^{p}_{MG}=\frac{\sigma\tilde{\alpha}_{p}}{1-\frac{1}{2}\sigma\tilde{\alpha}_{p}},
\end{equation}
so that the local field factor is
\begin{equation}\label{local2D}
\Re{\{\frac{\chi^{p}_{MG}}{\sigma\tilde{\alpha}_{p}}\}}=\frac{\Re{\{\epsilon^{p}_{MG}\}}+1}{2}.
\end{equation}
The dipole moment induced by an external field $\vec{E}^{\omega}_{0}(\vec{r})$ parallel to the plane on the emitter rod at $\vec{r}$ is
 $\vec{p}(\vec{r})=\epsilon_{0}\tilde{\alpha}_{p}l\vec{E}^{\omega}_{0}(\vec{r})$, $l$ being the length of the rod.
The Dyson propagator of transverse TE modes in the long-wavelength limit is,
\begin{equation}
G^{2D}_{\perp}(q,\omega)=\frac{-1}{\epsilon_{MG}^{p}k^{2}-q^{2}}.
\end{equation}
Correspondingly, the spectrum of coherent bulk modes reads,
\begin{eqnarray}\label{g2D}
\Im{\{\gamma_{\perp}^{2D}\}^{bulk}}&=&\Im{\{\int\frac{\textrm{d}^{2}q}{(2\pi)^{2}}G^{2D}_{\perp}(q,\omega)\}}\nonumber\\&=&
\frac{-1}{4}\Bigl[1+\pi^{-1}\arctan{\Bigl(\frac{\Im{\{\epsilon^{p}_{MG}\}}}{\Re{\{\epsilon^{p}_{MG}\}}}\Bigr)}\Bigr].
\end{eqnarray}
In function of Eqs.(\ref{local2D},\ref{g2D}),
the coherent power of TE modes is,
\begin{eqnarray}\label{wg}
l^{1}W^{Coh.}_{pMG}(\vec{r})&=&\frac{-\omega}{2l}\Im{\{\vec{p}(r)\cdot[\vec{E}^{\omega}_{0}(\vec{r})]^{*}\}}\\
&=&\frac{-\omega^{3}\epsilon_{0}}{4c^{2}}|\tilde{\alpha}_{p}|^{2}\Re{\{\frac{\chi^{p}_{MG}}{\sigma\tilde{\alpha}_{p}}\}}\Im{\{\gamma_{\perp}^{2D}\}^{bulk}}|\vec{E}^{\omega}_{0}(\vec{r})|^{2}.\nonumber
\end{eqnarray}
Substitution of Eqs.(\ref{local2D},\ref{g2D}) in the expression above yields Eq.(\ref{wcohMG2D}).

\end{document}